\documentstyle[fleqn,prb,aps,eqsecnum,amssymb,preprint]{revtex}

\begin{document}
\draft
\title{Magnetic order and disorder in the
frustrated quantum Heisenberg antiferromagnet in two dimensions}
\author{H. J. Schulz}
\address{
Laboratoire de Physique des Solides
\cite{assoc},
 Universit\'{e} Paris-Sud,
91405 Orsay,
France }
\author{T. A. L. Ziman and D. Poilblanc}
\address{
Laboratoire de Physique Quantique
\cite{assoc},
 Universit\'{e} Paul Sabatier,
31602 Toulouse,
France }
\maketitle
\begin{abstract}
We have performed a numerical investigation of the ground state properties
of the frustrated quantum Heisenberg antiferromagnet on the square lattice
(``$J_1-J_2$ model''), using exact diagonalization of finite clusters with
16, 20, 32, and 36 sites. Using a finite--size scaling analysis we obtain results for a number of
physical properties: magnetic order parameters, ground state energy, and
magnetic susceptibility (at $q=0$). For the unfrustrated case these results 
agree with series expansions and quantum Monte Carlo calculations 
to within a percent or better. In order to assess the reliability of
our calculations, we also investigate regions of parameter space with
well--established magnetic order, in particular the non--frustrated case
$J_2<0$. We find that in many cases, in particular for the intermediate
region $0.3 < J_2/J_1 < 0.7$, the 16 site cluster shows anomalous finite
size effects. Omitting this cluster from the analysis, our principal
result is that there is N\'eel type order for $J_2/J_1 < 0.34$ and collinear
magnetic order (wavevector $\bbox{Q}=(0,\pi)$) for $J_2/J_1 > 0.68$. There
thus is a region in parameter space without any form of magnetic order.
Including the 16 site cluster, or analyzing the independently calculated
magnetic susceptibility we arrive at the same conclusion, but with modified
values for the range of existence of the nonmagnetic region. We also find
numerical values for the spin--wave velocity
and the spin stiffness. The spin--wave velocity remains finite at
the magnetic--nonmagnetic transition, as expected from the nonlinear sigma
model analogy.
\end{abstract}


\section{Introduction}
In this paper  we consider a simple example of quantum frustrated
antiferromagnetism, namely the frustrated spin--1/2 Heisenberg model, 
with Hamiltonian
\begin{equation}
\label{ham}
    H = J_1 \sum_{\langle i,j \rangle} \bbox{S}_i \cdot \bbox{S}_j
  + J_2 \sum_{\langle i,j' \rangle} \bbox{S}_i \cdot \bbox{S}_{j'} \;\;.
\end{equation}
The spin operators obey $\bbox{S}_i \cdot \bbox{S}_i = 3/4$, and
 $J_1=1$ throughout this paper. The notations $\langle i,j \rangle$ 
and $\langle i,j' \rangle$ indicate summation over the 
nearest-- and next--nearest neighbor bonds on a square lattice, 
each bond being counted once.
While the model has attracted most
attention as  a simplified model\cite{inui_doniach_gabay} of the effects of 
doping on copper oxide planes in the high-temperature superconducting
copper oxides, it is of rather more general interest.
A complete understanding would provide a clear example
of answers to several general questions about
quantum phase transitions.
\par
The first question is that even in
a ground state with rather classical
looking symmetry, in this case an antiferromagnet, how do we
show unequivocally that the order really is of long range
and not simply local? How do we calculate
physically measurable correlations without relying on 
low order perturbation theory? In the present case, for small 
frustration  the appearance, 
in the limit of infinite size, of spontaneous
symmetry breaking is displayed in a relatively simple model. Indeed  the
renewed  interest in the model was because of doubts that
the unfrustrated case would display long--range order in
the thermodynamic limit. While such doubts are now relatively rare
thanks to  extensive numerical calculations and tighter
rigorous limits for higher spin and lower spin symmetry,\cite{kubo_kishi} 
there is as yet no rigorous
proof for  the isotropic spin one--half model in two spatial dimensions.
One reason for the present study is to test the quantitative  success of 
ideas of
finite size scaling as applied to numerical diagonalizations
that are perforce limited to what seem unhelpfully small samples.
\par The history of finite size effects goes back to Anderson in the
nineteen--fifties,\cite{anderson_52}
who first invoked the fact that the 
infinite degeneracy of the ground-state with spontaneously
broken continuous symmetry must be manifest in a large number 
of nearly degenerate
states in a large but finite system. 
This idea  of a ``tower" of states whose 
degeneracy corresponds to the ultimate symmetry,
and whose energy scales determine the 
long distance parameters of the spontaneously broken model of the 
infinite system has since been made more
precise and less dependent on perturbative
concepts in the language of non-linear sigma models.
\cite{hasenfratz_niedermayer} The model we  consider
here has the advantage over, for example, the triangular 
or Kagom\'e
antiferromagnets\cite{bernu_lhuillier_pierre,azaria_delamotte_mouhanna} in that
the classical limit has a simpler unit cell and thus the 
structure of the towers should be simpler to test. One of our aims here
will be to show that it is possible to extract the parameters of
the long wavelength physics in the ordered regime. 
In practice the difficulties of applying  finite
size studies are still considerable: there are 
subleading as well as leading corrections which make
the ultimate goal of reliable quantitative calculations
difficult even here. It is helpful that  we may easily
stabilize the ordered state to study the disappearance of order in a controlled
fashion by applying negative $J_2$.
\par A second general question relevant to other
quantum phase transitions, is whether the finite size methods
developed can be applied all the way to a critical point at which the 
order may disappear with a continuous transition. The first step is to identify the parameter 
$J_{2c}$
of this critical point
unequivocally; even its existence is still a matter for contention. Indeed
some self--consistent
spin-wave expansions have been interpreted as indicating 
a first order
transition\cite{takahashi_spinwave,nishimori_spinwave,mila_bruder}, at least for large spin.
We shall present results which we feel are rather convincing as to the
existence of a critical point and a reasonably accurate estimate of its 
value.
  
\par A third question, separate from the study of ordered antiferromagnetism,
is the question of what happens when this order disappears. In the 
mapping of quantum interacting ground states to 
thermodynamics of classical models in higher dimension,
there is at first sight a difference in that quantum phase transitions
tend to show order--order  rather than  order-disorder
transitions. Of course what one means by ``order"
is crucial to such a distinction. Here 
an ordered state would be understood to 
have  long range order in a different local order parameter,
for example a spin-Peierls dimerization variable or chirality
parameter. In this paper we do not discuss in detail the nature of
the intermediate state, but we do produce evidence that at least
it corresponds to one of zero uniform susceptibility. This is
compatible with either the chiral or dimerized phases favored 
previously.\cite{schulz_ziman} With increasing sample size both 
phases would have an 
exponentially vanishing ferromagnetic susceptibility.

We use exact finite--size diagonalization on clusters of $N = 16, 20, 32,
36$ sites (fig.\ref{f1}) with periodic boundary conditions. These are all
the clusters
accessible by our present calculational means that both respect the square
symmetry of the lattice and do not frustrate the collinear magnetic state
expected at large $J_2$ (this last condition is violated, for example, for the 18 and
26 site clusters, and more generally whenever $N$ is not an integer
multiple of 4). The present study is an improvement of our previous
finite-size calculation,\cite{schulz_ziman} which was restricted to the 
untilted four by
four
and six by six lattices. We shall see that the inclusion of the
intermediate sizes is extremely useful in allowing us to test the
consistency of the calculation. The final qualitative conclusions are
not drastically changed but estimates of critical parameters in
particular are altered. We are also able to articulate questions
about whether it is advisable to extrapolate from the sixteen site
cluster. In our previous study we had no choice even though the rather special
hypercubic property of the 16 site cluster was a concern. 

 As long as frustration is not too strong, the ground state of the model
(\ref{ham}) is expected to have long--range antiferromagnetic order, and
then the low--energy long--wavelength excitations are expected to be
described by the quantum nonlinear sigma model \cite{chakravarty_halperin},
with action \begin{equation}
S =  \frac{\rho_s}{2} \int d^2r \int_0^\beta d\tau \left[ (\nabla
\bbox{n})^2
+ \frac{1}{c^2}\left( \frac{\partial \bbox{n}}{\partial \tau} \right)^2
\right] \;\;.
\end{equation}
Here $\bbox{n} \equiv \bbox{n}(\bbox{r},t)$ is the local orientation of the
staggered magnetization, with $|\bbox{n}| = 1$, $c$ and $\rho_s$ are the
spin--wave velocity and spin stiffness, and the inverse temperature $\beta$ has
to be taken to infinity here as we are interested in ground state properties.
Lowest order spin--wave theory gives
\cite{chakravarty_halperin,einarsson_seff}$c_0 = \sqrt{2(1- 2J_2/J_1)}J_1$,
$\rho_{s0} = (J_1 - 2 J_2)/4$, but there are of course important quantum
fluctuation corrections to these quantities. One way to extract these
corrections  from finite size data will be discussed in sec. \ref{3b}
below. We note that the magnetic susceptibility at $\bbox{q}=0$ is given by
 $\chi = \rho_s/c^2$, which in spin--wave theory equals $1/(8J_1)$.
A major aim of the paper is to obtain controlled estimates of the different 
parameters beyond spin--wave theory. A summary of current results for 
the unfrustrated case can be found in recent review 
articles.\cite{manousakis_review,barnes_review}

\section{Numerical procedures and results}
We wish to find eigenvalues and eigenvectors of the Hamiltonian (\ref{ham})
on large clusters. In order to achieve this, and given that
computational
power is and will remain limited, it is necessary to use the symmetries
of
the problem to reduce the size of the corresponding Hilbert space as much as
possible. For the $N = 16,32,36$ clusters we use:
\begin{enumerate} \item
 translational symmetry ($N$
operations for an $N$--site cluster).
\item reflection on horizontal ($R_-$)
and vertical ($R_|$) axes (4 operations).
For the $N=16$ and $N=36$ cluster, both symmetry axis pass in between
rows of spins. However, for $N=32$, the $R_-$--axis coincides with the
central row of spin (see fig.\ref{f1}). 
\item if a given eigenstate has
the same eigenvalue under $R_-$ and $R_|$, then reflection on the diagonal
running from the lower left to the upper right of the cluster ($R_/$)
is also a symmetry operation, and can be used to further reduce the size of
the Hilbert space by a factor 2. For the 32 site cluster, this operation has
to be followed by a translation to remap the cluster onto itself.
\item if the z--component $S_z$ of the total
magnetization (which commutes with the Hamiltonian) is zero, then the spin
inversion operation $|\uparrow\rangle \leftrightarrow |\downarrow\rangle$
is also a symmetry and leads to a further reduction by a factor 2. In
principle a further considerable reduction of the Hilbert space could be
achieved by using the conservation of the total spin $\bbox{S}^2$. However,
there does not seem to be any simple way to efficiently incorporate this
symmetry. Note that we could equally have chosen reflection operators
centered on a particular site.
\end{enumerate}
The point group operations $Id, R_-, R_|, R_/$ generate the point group symmetry
$C_{4v}$. These operations are only compatible with the translational
symmetry for states of momentum $\bbox{Q} = 0$ or $\bbox{Q} = (\pi, \pi)$.
In particular, for our clusters the
ground state is always at $\bbox{Q} = 0$.
For the 20 site clusters reflections are not symmetry operations, and we
use rather a rotation by $\pi/2$ as generator of the point group.
 The symmetry group at the interesting momenta $\bbox{Q} =
0$ or $\bbox{Q} = (\pi, \pi)$ then is $C_4$.

We use a basis set characterized by the value of $S_{zi}$ at each lattice
site $i$.
An up (down) spin is represented by a bit 1 (0) in a computer word. Thus, a
typical spin configuration (e.g. for a linear  system of 4 spins) would be
represented as
\begin{equation} \label{st1}
|\uparrow \uparrow \downarrow \uparrow \rangle = 1101_2 = 13
\end{equation}
To implement the symmetry, we do not work in this basis, but use rather
symmetry--adapted basis states. E.g. to remain in the one--dimensional toy
example, instead of (\ref{st1}) we use the normalized basis state
\begin{equation}    \label{st2}
\frac12 (|\uparrow \uparrow \downarrow \uparrow \rangle
+ |\uparrow \uparrow \uparrow \downarrow \rangle
+ |\downarrow \uparrow \uparrow \uparrow \rangle
+ |\uparrow \downarrow \uparrow \uparrow \rangle )
= \frac12 (|13\rangle + | 14 \rangle + | 7\rangle + | 11 \rangle )
\equiv |7)
\end{equation}
where the lowest ({\em ''minimal''}) integer of the 4 states occurring in
(\ref{st2}) is used to represent the state.

Our procedure to determine eigenvectors and eigenvalues procedes in three
steps: (i) starting from an arbitrary basis state of given symmetry and
$S_z$, the whole Hilbert space is generated by repeated application of the
$J_1$ part of the Hamiltonian, and the basis set is stored; (ii) the
Hamiltonian matrix is calculated and stored in two pieces, corresponding to
the $J_1$ and $J_2$ parts of the Hamiltonian; (iii) the matrix is used in a
Lanczos algorithm to obtain eigenvalues and eigenvectors of the Hamiltonian.
The principal difficulty in steps (i) and (ii) is that application of the
Hamiltonian to a state represented by a ``minimal'' integer will of course
in general not produce another minimal integer state, but rather a state
that needs to be brought into minimal form by the application of a symmetry
operation. The trivial solution would be to try out all possible operations.
This however would be extremely time consuming (there are 576 symmetry
operations for the 36 site cluster!). Instead we use a different procedure
\cite{pierre}: the basis states are coded in a computer word so that the
$R_|$ operation corresponds to the exchange of the two halfwords. Each
halfword then can be an integer between $0$ and $2^{N/2}$. We then
create a list specifying for each {\em halfword} the corresponding
minimal state (integer) and the symmetry operation(s) connecting them.
The length of this list is relatively moderate ($2^{18}$ at worst),
and it
can be easily kept in computer memory.

The minimal state corresponding
to a given basis state is now determined by looking in this list for the
minimal states corresponding to the two halfwords. If necessary, the
two halfwords are exchanged (i.e. a $R_|$ operation is performed), so
that the smallest of the two halfwords constitutes the high--bit
halfword of the resulting state. Finally, the symmetry operation
leading to this high-bit halfword is applied to the remaining halfword.
In about 80\% of the cases this symmetry operation is uniquely
determined. In the remaining cases, more than one symmetry operation has
to be tried out in order to find the minimal state. However, the extra
calculational effort is relatively small: e.g. for $N=36$, only for
about a thousand out of $2^{18}$ possibilities are there more than eight
symmetry operations to be tried out.
Using this method for $N=36$ the CPU time needed to calculate the basis
set and the Hamiltonian matrix is approximately 30 min and 90 min.,
respectively.\cite{rem2} For the smaller clusters, CPU time requirements
are obviously much less. In table I we show the size of the basis set and
the number of non--zero matrix elements for states of $A_1$
($N=16,32,36$) or $A$ ($N=20$) symmetry at $\bbox{Q} = 0$. These
are the subspaces containing the groundstate, apart from the case of
relatively large $J_2$ on the $N=20, 36$ clusters,  where the ground state
has point group symmetry $B$ ($N=20$) or $B_1$ ($N=36$). Note that the
number of basis states for the larger clusters is very close to the
naive expectation $\left(
\begin{array}{c} N \\ N/2 \end{array}
  \right)/(16N)$ ( $\approx
1.57554 \times 10^7$ for $N=36$).

The number of matrix elements e.g. for $N=36$ is still enormous. It is
however obvious that the matrix is extremely sparse: on the average, there
are fewer than 80 nonzero elements per line, which has in all $15,804,956$
positions. One obviously only wants to store the addresses and values
of the nonzero matrix elements. This would still need two computer words
per non--zero matrix element, however, this requirement can be further
reduced noting that all matrix elements are of the form $H_{i,j} =
J_{1,2}(\lambda_i/\lambda_j) I_{i,j}$, where the $\lambda_i$ are
the normalization factors of the symmetrized basis states (like the
factor $1/2$ in (\ref{st2})), and the $I_{i,j}$ are small integers,
which in the vast majority of cases equal unity. More specifically,
$I_{i,j}$ is the number of times the action of the Hamiltonian on an
unsymmetrized basis state $|i\rangle$ creates another
unsymmetrized basis state $|j\rangle$ or a state related to $|j\rangle$ by
a symmetry operation. The
values of the $\lambda_i$ intervening in a given matrix element can be easily
determined during the calculation, and we thus just store the positions
of the unit integer. For the cases where $I_{i,j} = n \neq 1$, the
corresponding position is stored $n$ times. Finally, a Cray computer
word has
64 bits, and therefore can accomodate two addresses. In this way the
whole matrix for the 36 site cluster can be stored in about 5 gigabytes,
which is relatively easily available as disk space at the computing
facility we are using. Space requirements could be further reduced by
a factor 2
using the symmetry of the Hamiltonian matrix, however, this would have
lead to a rather important loss of speed in the subsequent matrix
diagonalization.

\begin{figure}
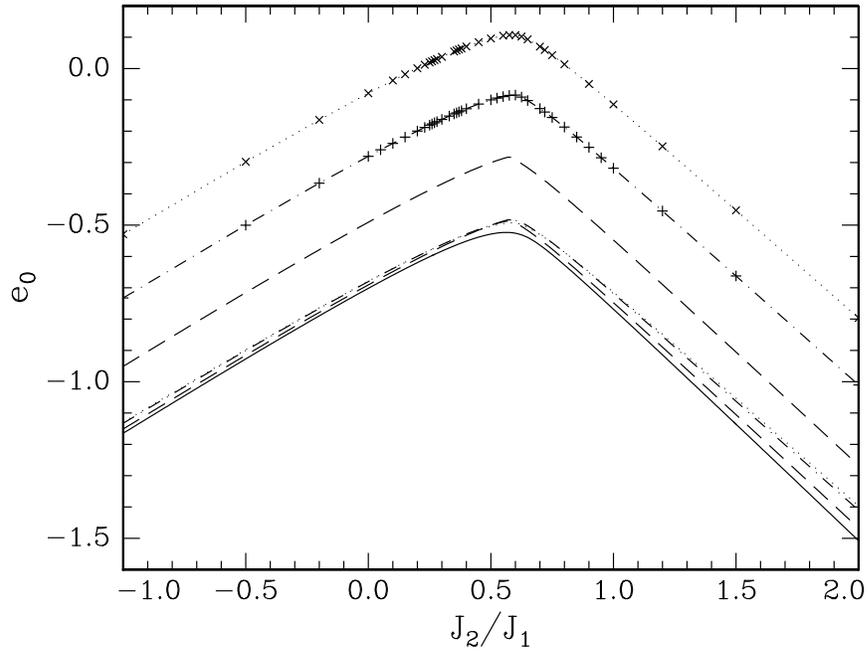

\caption{The ground state energy per site as a function of $J_2/J_1$ for
$N=16$ (full line), $N=20$ (dashed line), $N=32$ (dash--dotted line), and
$N=36$ (dotted line). For clarity, the curves for $N=20,32,36$ are also
displayed shifted upwards by $0.2$, $0.4$, and $0.6$, respectively. For
$N=16, 20$ we have results for $J_2/J_1$ in steps of $0.01$, and only a
continuous curve is displayed. For $N=32,36$, we have only results at the
points indicated, and lines are a guide to the eye.}
 \label{f2}
\end{figure}

To obtain the groundstate eigenvalue and eigenvector of the Hamiltonian we
use the standard Lanczos algorithm, implemented by the Harwell library
routine EA15AD. This routine performs rather extensive convergence checks
and we thus avoid to perform unnecessary time--consuming Lanczos iterations.
The main problem at this level is the use of the still rather large matrix
($\approx 5$ gigabytes). The matrix clearly does not fit into the main memory
of a Cray--2 (2 gigabytes). We therefore store the matrix on disk, and read
it in by relatively small pieces, whenever a new piece is needed. This
operation can be made
computationally efficient by using ``asynchronous'' input operations, which
allow one to perform calculations in parallel with the read--in operation
for the next piece of the matrix. Moreover, using more than one input
channel simultaneously the read--in operation can be further accelerated.
In this way the total time overhang due to the continuous read--in of the
matrix can be kept below 20\% of total CPU time. To reach a relative
accuracy of $10^{-6}$ for $N=36$, we need between 40 min. ($J_2 = 0$) and 3
hours ($J_2/J_1 \approx 0.6$, slowest convergence) CPU time.
We have performed a number of checks to insure the correctness of the
numerical algorithm. The most important one is to calculate the groundstate
energy for {\em ferromagnetic} interaction ($J_{1,2} < 0$), which of course
is known to be $(J_1 + J_2)N/2$. However, the numerical calculation in the
$S_z=0$ subspace is nontrivial because the Hilbert space and, up to an
overall minus sign, the matrix are of course the same as for the
antiferromagnetic case. We also compared our results with previous
finite size calculations
\cite{dagotto_moreo,figueirido_16site,lin_spins,poilblanc_16,poilblanc_dagotto}
(for $N=16, 20, 32$) and quantum Monte Carlo
results \cite{runge_mag} (for $N=36, J_2 = 0$), and found agreement in all
cases. Finally,
an independent check of the numerical accuracy of the Lanczos algorithm is
provided by starting the Lanczos iterations with different initial vectors.
In each case we found a relative accuracy of at least $10^{-6}$ for the
ground state eigenvalues. Similarly, expectation values calculated with the
eigenvector are found to have relative accuracy of $10^{-4}$. In table
\ref{t2} we list ground state energies of the different clusters for a number
of values of $J_2$. A more complete set of results is displayed in
fig.\ref{f2}.

More important for the following analysis are the values of
the $\bbox{Q}$--dependent magnetic susceptibility (or squared order parameter)
\begin{equation} \label{mq}
M_N^2(\bbox{Q}) = \frac{1}{N(N+2)} \sum_{i,j} \langle \bbox{S}_i \cdot \bbox{S}_j
\rangle e^{i \bbox{Q} \cdot (\bbox{R}_i - \bbox{R}_j)} \;\;.
\end{equation}
Following arguments by Bernu et al.\cite{bernu_lhuillier_pierre} we use a
normalization by a prefactor $\frac{1}{N(N+2)}$ instead of the usual $1/N^2$.
In the thermodynamic limit, these possibilities are obviously equivalent. 
However, for the relatively small cluster we are using, there are sizeable 
differences in the results of the finite--size scaling analysis. The choice in
eq.(\ref{mq}) is essentially motivated by the fact that in a perfect 
N\'eel state $M_N^2(\bbox{Q})$ is entirely 
size--independent.\cite{thanks_claire} More generally, this choice eliminates to a certain extent the overly strong contributions from the terms with $i=j$ in
eq.(\ref{mq}). 

Some values of $M_N^2(\bbox{Q})$ are shown in table \ref{t3}, and complete curves are in
fig.\ref{f3}. The values displayed (and used in the following analysis)
are always expectation values in the {\em true ground state}, e.g. for
large $J_2$ states of symmetry $B$ ($N=20$) or $B_1$ ($N=36$) are used.
From the results shown it is quite obvious that the dominant
type of magnetic order changes from $\bbox{Q} = (\pi,\pi)$ at relatively small
$J_2$ (N\'eel state) to $\bbox{Q} = (\pi,0)$ at larger $J_2$ (collinear
state). How exactly this change occurs will be clarified in the following
section.

 \section{Finite--size scaling analysis}
\subsection{Order parameters}
\begin{figure}
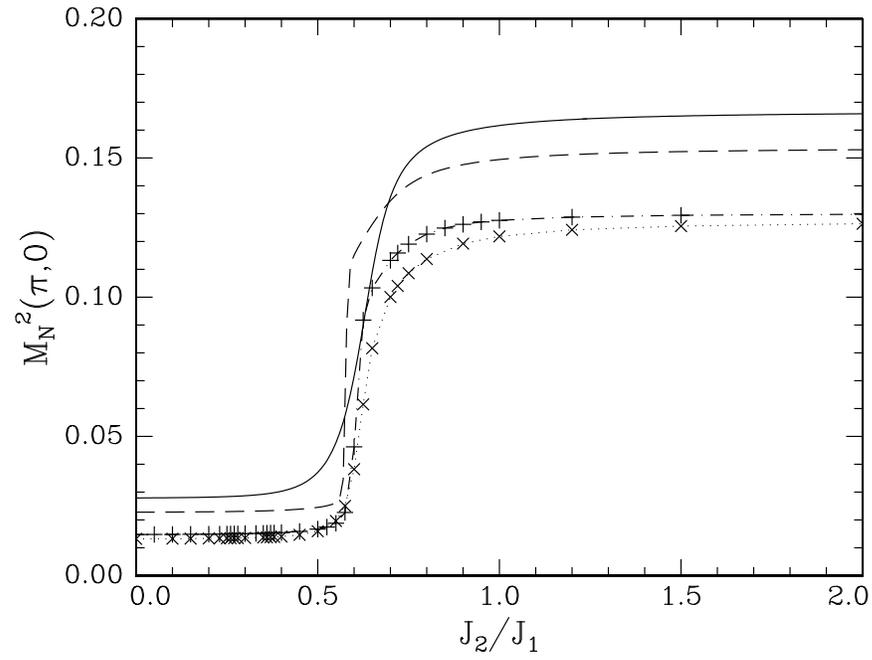

\caption{The magnetic susceptibility $M(\protect\bbox{Q})$ at (a) $\protect\bbox{Q}=(\pi,\pi)$
and (b) $\protect\bbox{Q}=(\pi,0)$. The symbols and linetypes are the same as in
fig.\protect{\ref{f2}}. }
\label{f3}
\end{figure}

The results shown in fig.\ref{f3} show a transition between a N\'eel ordered
region for $J_2 \lesssim 0.5$ to a state with so--called collinear order
(i.e. ordering wavevector $\bbox{Q} = (\pi,0)$) at $J_2 \gtrsim 0.6$. To
analyze the way this transition occurs in more detail, we use finite--size
scaling arguments.\cite{neuberger_ziman} In particular, it is by  now well
established that
the low--energy excitations in a N\'eel ordered state are well described by the
nonlinear sigma model. From this one can then derive the finite--size
properties of various physical quantities. The quantity of primary interest
here is the staggered magnetization $m_0(\bbox{Q_0})$ defined by
\begin{equation}
m_0(\bbox{Q}_0) = 2 \lim_{N \rightarrow \infty} M_N(\bbox{Q}_0) \;\;,
\end{equation}
where $\bbox{Q}_0 = (\pi,\pi)$. The normalization is chosen so that
$m_0(\bbox{Q}_0)=1$ in a perfect N\'eel state. The leading finite size
corrections to $m_0$ are given by
\begin{eqnarray}
\nonumber
M_N^2(\bbox{Q}_0) & = & \frac14 m_0(\bbox{Q}_0)^2 + 1.2416
\frac{\kappa_1^2}{\sqrt{N}} + ... \\
& = & \frac14 m_0(\bbox{Q}_0)^2 (1 + \frac{0.6208c}{\rho_s \sqrt{N}}
 + ... ) \;\;,
\label{fsm}
\end{eqnarray}
where for the infinite system $\kappa_1$ gives the amplitude of the
diverging matrix element of the spin operator between the ground state and
 single magnon states at $\bbox{Q} \approx \bbox{Q}_0$.

\begin{figure}
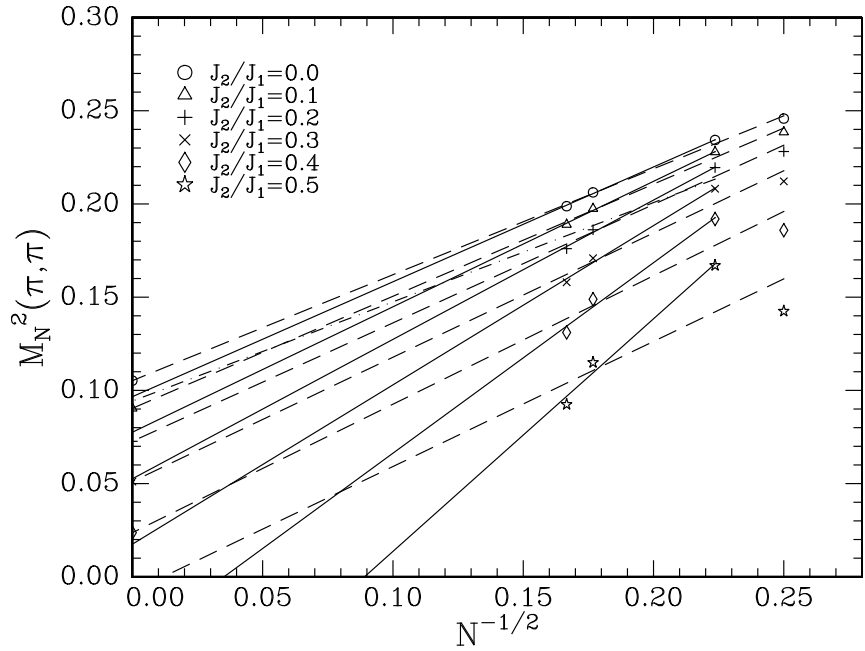

\caption{Finite size results for $M_N^2(\protect{\bbox{Q}}_0)$ for different
values of $J_2$. The dashed lines are least squares fits to the data
according to eq.(3.2), using all available clusters. The full lines
are fits using only $N=20,32,36$. The dash--dotted line is the leading
finite size behavior expected at $J_2=0$ (see eq.(3.2)).}
\label{f4}
\end{figure}

\begin{figure}
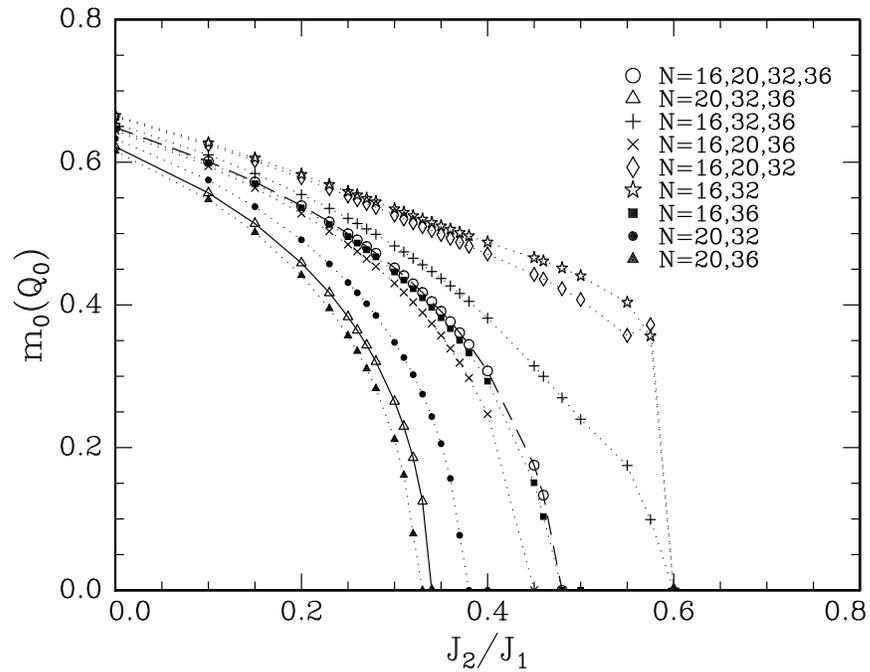

\caption{The staggered magnetization $m_0(\protect{\bbox{Q}}_0)$ as a
function of $J_2/J_1$ using different combinations of clusters (a). In (b)
the ``critical'' region $J_2>0$ is shown enlarged.}
\label{f5}
\end{figure}

Least square fits of our finite--size results to eq.(\ref{fsm}) are shown in
fig.\ref{f4}. For small values of $J_2$ the scaling law is quite well
satisfied: e.g. for $J_2=0$ the four data points in fig.\ref{f4} very nearly
lie on the ideal straight line, and the extrapolated value of the staggered
magnetization, $m_0(\bbox{Q}_0)=0.649$, is quite close to the best
current estimates, $m_0(\bbox{Q}_0)=0.615$.\cite{com_m0}
Using the same type of finite
size extrapolations for other values of $J_2$, we obtain the results
indicated by a dashed line in fig.\ref{f5}~.

For $J_2 = 0$, a check on the reliability of our method can be obtained by
comparing the numerical results with what one would expect from
eq.(\ref{fsm}), using the rather reliable results for $m_0$, $c$,
and $\rho_s$ obtained by series expansion 
techniques.\cite{singh_series,singh_huse,weihong_series} 
The curve expected from eq.(\ref{fsm})
is shown as a dash--dotted line in fig.\ref{f4}. It appears that there are
sizeable but not prohibitively large next--to--leading corrections.

Another measure of the reliability of the finite--size extrapolation
can be obtained
comparing results obtained by the use of different groups of clusters. For
negative $J_2$, i.e. {\em
nonfrustrating interaction}, the values of $m_0(\bbox{Q}_0)$ are nearly
independent
of the clusters sizes used, and the results in fig.\ref{f5} therefore are
expected to be quite accurate. In this region the next nearest neighbor
interaction stabilizes the antiferromagnetic order and therefore
the
staggered magnetization tends to its saturation value unity for large
negative $J_2$.
On the other hand, for positive $J_2$ the interaction is frustrating. In
this case, the agreement between different extrapolations is less good. We
note however, that in all but two cases the staggered magnetization tends to
zero as in a second order phase transition, with a critical value of $J_2$
between $0.34$ and $0.6$. The question than arises as to which extrapolation
to trust most. In fact, none of the clusters considered here is free of
some peculiarity: for $N=16$ and $J_2=0$, there is an extra symmetry, because
with only nearest neighbor interactions
this cluster is in fact equivalent to a $2 \times 2 \times 2 \times 2$
cluster on a four--dimensional hypercubic lattice; the $N=20$ cluster
has a lower symmetry than all the others ($C_4$ instead of $C_{4v}$);
for $N=20$ and $N=36$ the
ground state changes symmetry with increasing $J_2$; finally the 20 and 32
site clusters are unusual in that they are rotated, by different angles,
with respect to the lattice directions. A priori, one might then argue that the best choice should be the least  biased one, including all available clusters.
As
indicated by the dashed line in fig.\ref{f5}, this leads to a critical value of
$J_2$ for the disappearance of antiferromagnetic order of $J_{2c} \approx
0.48$.

However, from fig.\ref{f4} it is quite clear that for $J_2 \ge 0.35$ the 16 site cluster is highly anomalous in that
$M_N^2(\bbox{Q}_0)$ increases going to the next bigger cluster, whereas in
all other cases there is a decrease with increasing size. Clearly, in
fig.\ref{f4} a much better fit is obtained in this region by omitting the
$N=16$ results, leading to a reduced value, $J_{2c} \approx 0.34$ as
indicated by the full line in fig.\ref{f5}. The anomalous results
obtained from the $N=16,20,32$, $N=16,32,36$ and $N=16,32$ fits are 
certainly  due to an over--emphasis put onto the $N=16$ results. Similar
anomalous behavior of the 16 site cluster occurs in many cases in the region $0.3 < J_2 < 0.8$, and we therefore consider the results obtained using only
$N=20,32,36$ as more reliable. In particular, in this way we find a staggered magnetization of $0.622$ at $J_2=0$, only about one percent higher than the best current estimate, $m_0(\bbox{Q}_0) = 0.615$.
Beyond the precise value of the critical value $J_{2c}$ at which antiferromagnetic order disappears, the important result here, obtained by
the majority of fits, is the existence of a second order transition, located
in the interval $0.3 \le J_2 \le 0.5$. 

\begin{figure}
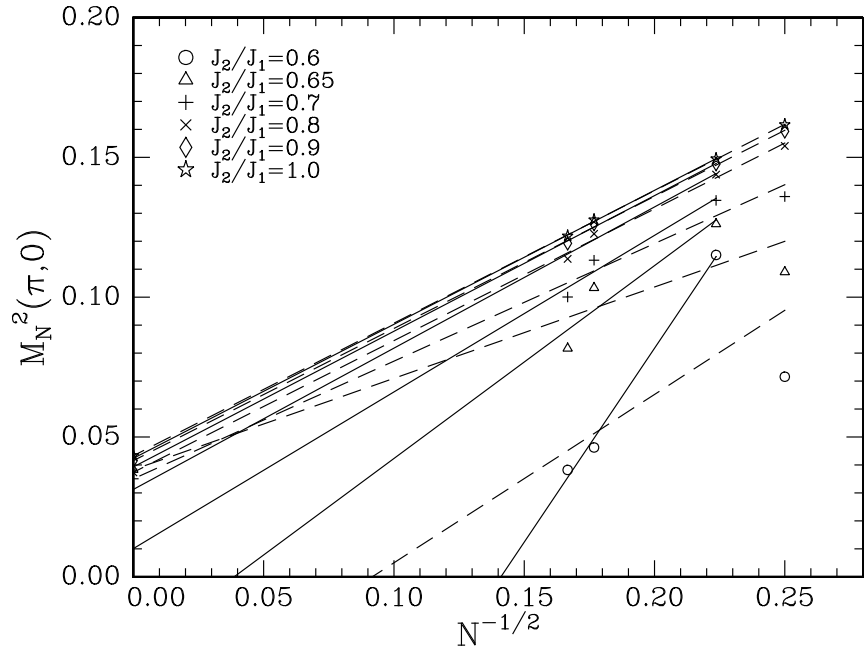

\caption{Finite size results for $M_N^2(\protect{\bbox{Q}}_1)$ for
different
values of $J_2$. The dashed lines are least squares fits to the data
according to eq.(3.3), using all available clusters. The full lines
are fits using only $N=20,32,36$.}
\label{f6}
\end{figure}

One might of course argue that it is not the $N=16$ but rather the $N=20$
cluster that is anomalous. However, closer inspection of the data in
figs.\ref{f4} and \ref{f6} clearly shows that the $N=20,32,36$ data points
remain reasonably well aligned even in the intermediate region $0.3\le
J_2 \le 0.8$, whereas the alignment for $N=16,32,36$ is much worse.
In the following, we will therefore mostly rely on the $N=20,32,36$
extrapolations.

We now follow the same logic to analyze the behavior for larger $J_2$,
where fig.\ref{f3} suggest the existence of magnetic order with ordering
wavevector $\bbox{Q}_1 = (\pi,0)$. Of course, this state again breaks the
continuous spin rotation invariance, and therefore the low energy
excitations are described by a (possibly anisotropic) nonlinear sigma model.
There is an additional breaking of the discrete lattice rotation symmetry
(ordering wavevector $(0,\pi)$ is equally possible), however, this does not
change the character of the low--lying excitations. The finite size
behavior is entirely determined by the low energy properties, and
therefore we expect a finite size formula analogous to eq.(\ref{fsm}):
\begin{equation}
\label{fsm1}
M_N^2(\bbox{Q}_1) = \frac18 m_0(\bbox{Q}_1)^2 +
\frac{const.}{\sqrt{N}} + ... \;\;.
\end{equation}
Here the factor $1/8$ (instead of $1/4$ in (\ref{fsm})) is due to the extra
discrete symmetry breaking which implies that finite--size ground states are
linear combinations of a larger number of basis states. Moreover, the linear
sigma model is anisotropic, because of the 
spontaneous discrete symmetry breaking of the ordering vector, and consequently
a precise determination of the
coefficient of the $\sqrt{N}$--term is not straightforward. The important
point here is however the $N$--dependence of the correction term in
eq.(\ref{fsm1}).

Least square fits of our numerical results to eq.(\ref{fsm1}) are shown
in
fig.\ref{f6}, and the extrapolated collinear magnetization
$m_0(\bbox{Q}_1)$ is shown in fig.\ref{f7}. For
$J_2 \ge
0.8$ eq.(\ref{fsm1}) provides a satisfactory fit to our data, even
though not quite as good as in the region $J_2 \le 0$ in the staggered
case, as shown by the spread of different fits in fig.\ref{f7} (compare
fig.\ref{f5} in the region $J_2 \le 0$). For smaller $J_2$ there is a
wide spread in the extrapolated results, depending on the clusters used.
We notice however that for the majority of clusters used, there is a
common feature: $m_0(\bbox{Q}_1)$ remains finite down to $J_2 = 0.65$,
and then suddenly drops to zero at $J_2=0.6$. This would indicate a {\em
first order transition} to the collinear state somewhere in the interval
$0.6 < J_{2c} < 0.65$. This interpretation also seems consistent with
the raw data fo fig.\ref{f3}: the increase of $M_N^2(\bbox{Q}_1)$ around
$J_2 = 0.6$ is much steeper than the growth of $M_N^2(\bbox{Q}_0)$ with
decreasing $J_2$. From the $N=16,20,32,36$ extrapolation 
one then obtains a collinear magnetization which is roughly constant
above $J_{2c}$ at $m_0(\bbox{Q}_1) \approx 0.6$. Notice that the first--order
character of the transition is {\em not} due to the level crossings occurring
in the $N=20$ and $N=36$ clusters: if these clusters are omitted from the
extrapolation, the first order character is in fact strongest (cf.
fig.\ref{f7}).

On the other hand, inclusion or not of the $N=16$ cluster plays an important
role because this cluster shows again anomalous
behavior in the region of intermediate $J_2$: for $J_2 < 0.7$
$M_N^2(\bbox{Q}_1)$ increases when $N$ increases from $16$ to $20$,
contrary to what eq.(\ref{fsm1}) suggests. If one therefore omits the
$N=16$ cluster from the extrapolation, results quite consistent with a
second order transition are obtained, however this time with a critical
coupling $J_{2c} \approx 0.68$.

\begin{figure}
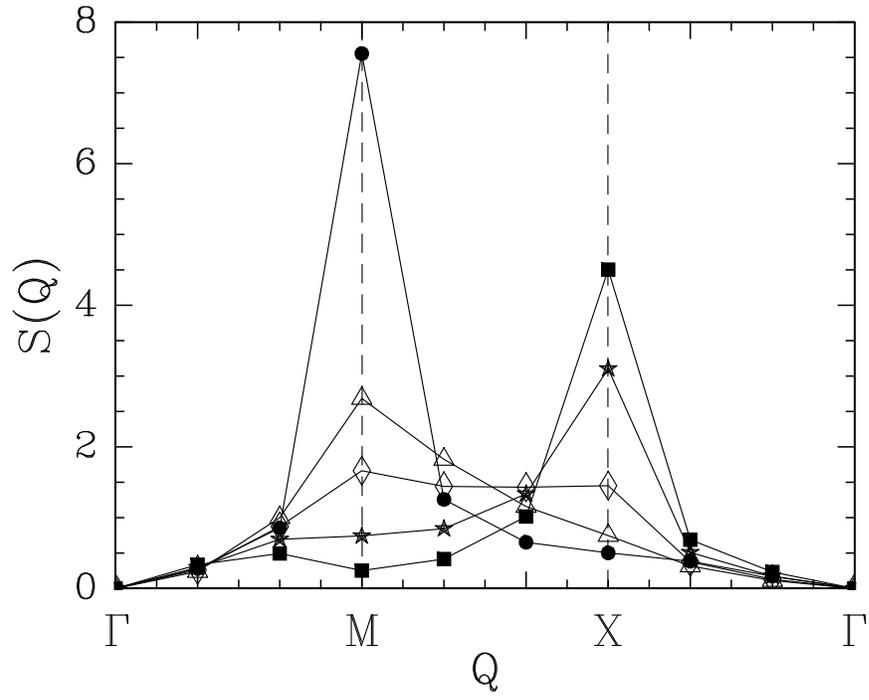

\caption{Magnetic structure factor, as obtained from the $N=36$
cluster, in the Brillouin zone for $J_2/J_1$=0
($\bullet$), 0.55 ($\bigtriangleup $), 0.6 ($\diamond$), 0.65
($\star$), 1 ($\blacksquare$). The points $\Gamma$, M, X are
$\protect{\bbox{Q}}=0,\protect{\bbox{Q}}_0,\protect{\bbox{Q}}_1$,
respectively.
Note that nowhere there is a maximum at a point different from M or X.
}
\label{f7a}
\end{figure}

Beyond quantitative results, the most important conclusion of this
analysis is the existence of a finite interval without magnetic long
range order: if all available clusters are included in the analysis,
this interval is $0.48 \lesssim J_2 \lesssim 0.6$. if, because of the
anomalies
discussed above one omits the $N=16$ cluster, the nonmagnetic interval
is increased to $0.34 \lesssim J_2 \lesssim 0.68$. The study of the ground state symmetry in
this region requires a detailed analysis of a number of different
non--magnetic order parameters and will be reported in a subsequent paper.
However, at this stage, the magnetic structure factor $S(\bbox{Q}) = (N+2)
M_N^2(\bbox{Q})$ already gives some valuable information: in fact, as shown
in fig.\ref{f7a}, with increasing $J_2$ the collinear peak at the X point
grows and the N\'eel peak at the M point shrinks, however there never is a
maximum at other points. There is thus no evidence for incommensurate
magnetic order.

\subsection{Ground state energy, spin--wave velocity, and stiffness constant}
\label{3b}
\begin{figure}
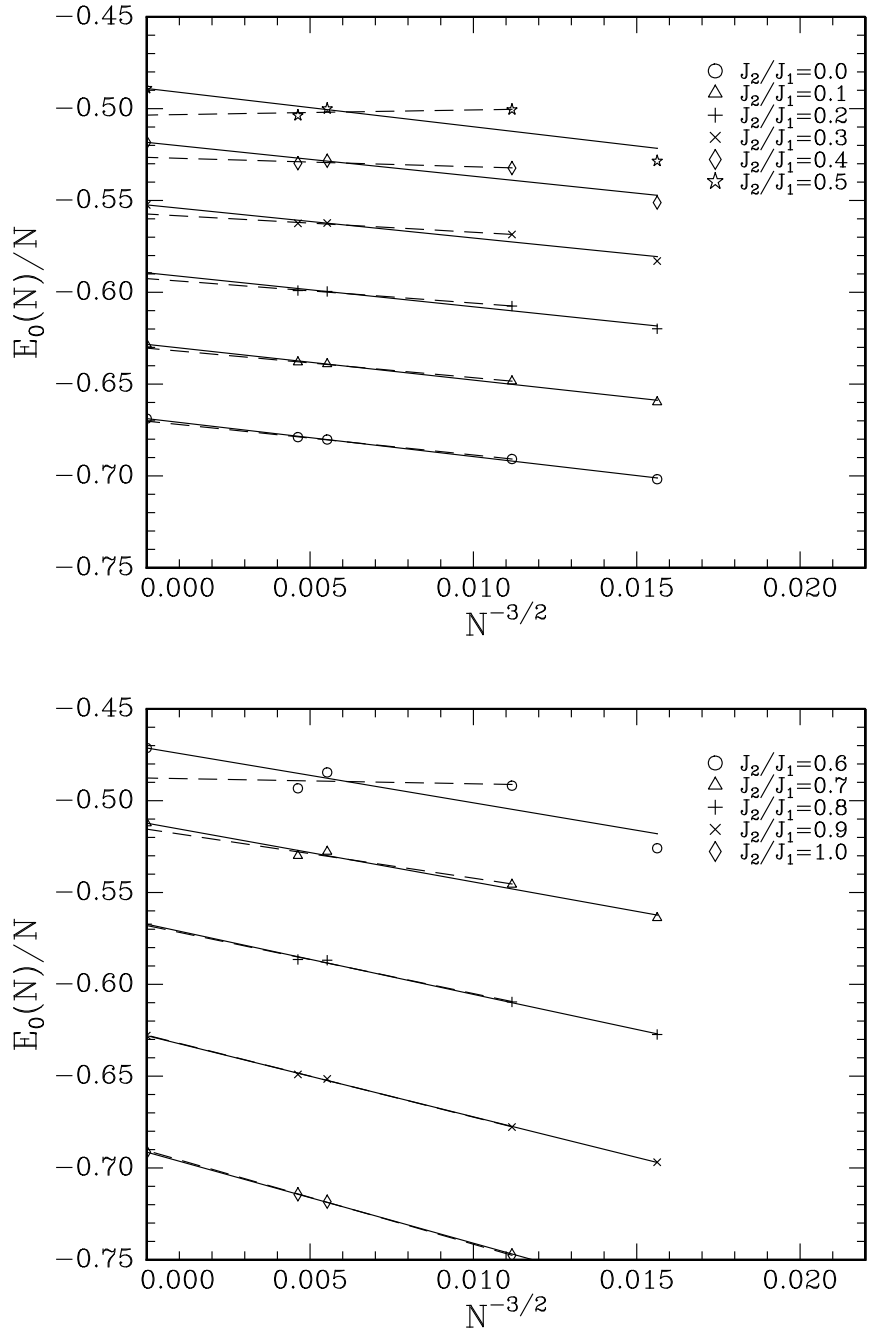

\caption{Finite size results for the ground
state energy per site for different
values of $J_2$. The full lines are least squares fits to the data according
to eq.(3.4), using all available clusters. The dashed lines
are fits using only $N=20,32,36$.}
\label{f8}
\end{figure}

\begin{figure}
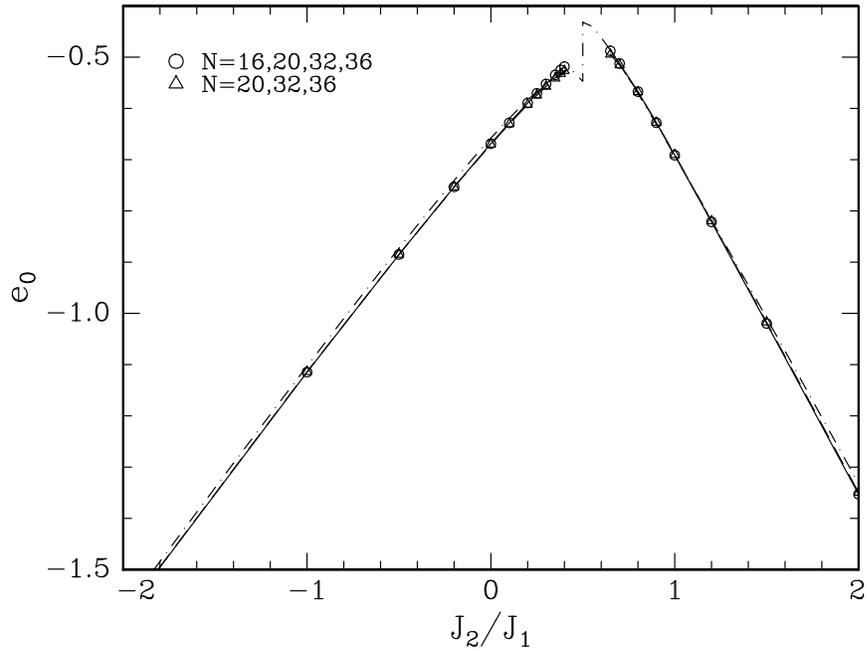

\caption{Ground state energy per site as obtained from finite size
extrapolation using eq.(3.4). In the intermediate region
$0.4<J_2< 0.65$ the extrapolation can not be used reliably, and no results
are shown. Results obtained using different clusters are undistinguishable
on the scale of this figure. The dash--dotted line is the spin--wave
result, eqs.(3.11) and (3.12).}
\label{f9}
\end{figure}

The ground state energy per site in the thermodynamic limit can be obtained
from the finite--size formula for an antiferromagnet
\cite{neuberger_ziman}
\begin{equation}
\label{fse}
E_0(N) / N = e_0 - 1.4372 \frac{c}{N^{3/2}} + ... \;\;,
\end{equation}
where $c$ is the spin--wave velocity. Again, in the collinear state, an
analogous formula holds, but with $c$ replaced by some anisotropy--averaged
value. Fits of our numerical results are shown in fig.\ref{f8}. Away from
the ``critical'' intermediate region, i.e. for $J_2 \le 0.2$ and $J_2 \ge
0.8$, eq.(\ref{fse}) provides a rather satisfying description of the results,
in particular if the $N=16$ cluster is disregarded. The fit is even
considerably better than that for the order parameters (compare
fig.\ref{f4}). This is certainly in large part due to the much weaker
finite size correction to the ground state energy, as compared to those
for the order parameters. On the other hand,
in the intermediate region $0.4 \le J_2 \le 0.7$, the fits are not very good.
In this region the ground state energy per site is rather irregular, for
example there is generally a {\em decrease} from $N=32$ to $N=36$ contrary
to what eq.(\ref{fse}) suggests. The failure of eq.(\ref{fse}) in the
intermediate region is of course not surprising, as the analysis of the
previous section showed the absence of magnetic order, which implies the
non--existence of an effective nonlinear sigma model and therefore the
invalidity of the extrapolation formula (\ref{fse}). The result of our
extrapolations is shown in fig.\ref{f9}. Over most of the region shown,
results from extrapolations using different clusters are indistinguishable
on the scale of the figure. Only close to the critical region is there a
spread of about 2 percent in the results. In particular, at $J_2=0$ we find
values between $e_0 = -0.668$ and $e_0=-0.670$, very close to the probably
best currently available estimate, obtained from large--scale quantum Monte
Carlo calculations, of $e_0 = -0.66934$.\cite{runge_extrapol}

\begin{figure}
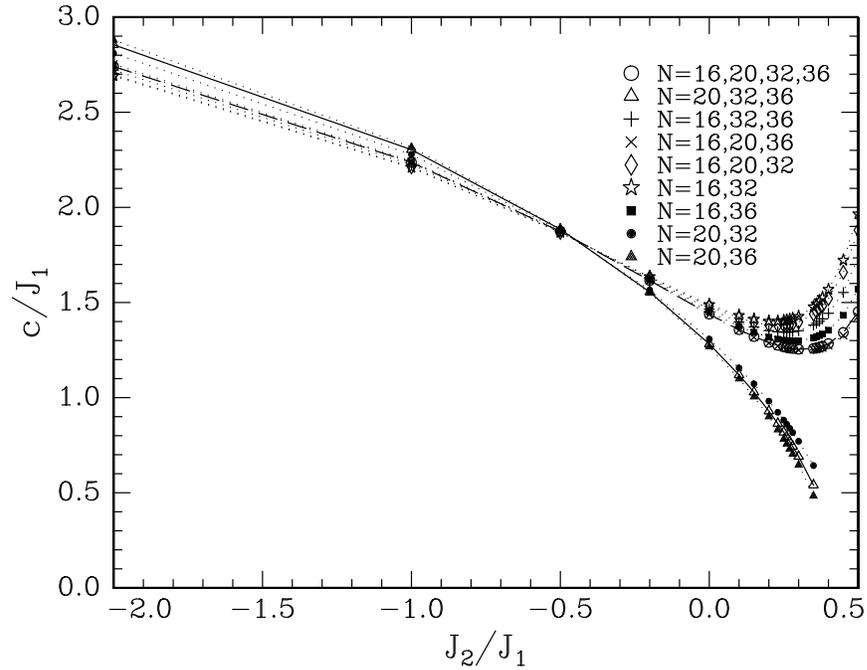

\caption{The spin wave velocity in the antiferromagnetic state
as obtained from finite
size extrapolation using eq.(3.4). No results are shown in
the region where according to the previous analysis there is no
antiferromagnetic order ($J_2 > 0.48$ or $J_2 > 0.34$ according to whether
the $N=16$ cluster is included or not).}
\label{f10}
\end{figure}

The amplitude of the leading correction term in eq.(\ref{fse}) allows for a
determination of the spin--wave velocity $c$. Results are shown in
fig.\ref{f10}. In this case, there is a wider spread in results. This is
certainly not surprising, given that this quantity is derived from the
correction term in eq.(\ref{fse}). Nevertheless, the agreement between
different extrapolations is reasonable for $J_2 \le 0$. At $J_2=0$ and
using all clusters we
find $c = 1.44 J_1$, close to but somewhat lower than the best spin--wave result
$c_{SW} = 1.65 J_1$. A smaller value is found from the $N=20,32,36$ extrapolation: $c=1.28$. For positive $J_2$ the
extrapolations give different answers, according to whether the $N=16$
cluster is included or not. This of course is due to the anomalous behavior
of this cluster in the energy extrapolations (see fig.\ref{f8}).
An important point should however be noticed: independently of the inclusion
of the $N=16$ cluster, at the critical value $J_{2c}$
for the disappearance of the antiferromagnetic order the spin--wave velocity
remains finite.

\begin{figure}
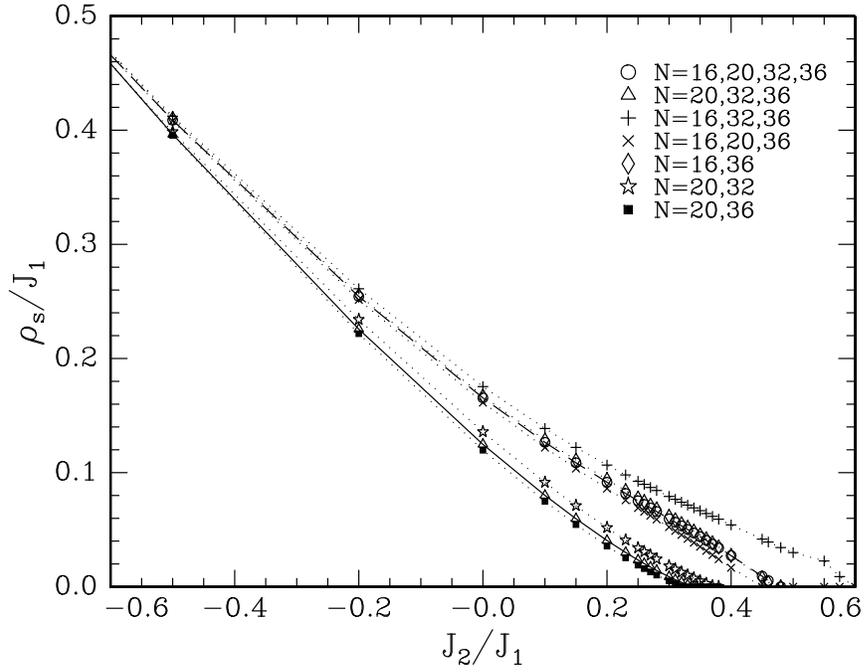

\caption{The spin stiffness in the antiferromagnetic state
as obtained from finite
size extrapolation using eq.(3.5). Lines are a guide to the
eye.}
\label{f11}
\end{figure}

The final parameter in the nonlinear sigma model is the spin stiffness
constant $\rho_s$. It can be found from our finite size results 
\cite{neuberger_ziman}
\begin{equation}  \label{fsr}
\rho_s = \frac{m_0(\bbox{Q}_0)^2 c}{8 \kappa_1^2} \;\;,
\end{equation}
with $\kappa_1$ determined from eq.(\ref{fsm}).\cite{erratum}
This relation determines the second form of eq.(\ref{fsm}) above.  
Results are shown in
fig.\ref{f11}. Again, for the same reasons as before, there is some scatter
in the results, because of the use of the correction terms in
eqs.(\ref{fsm}) and (\ref{fse}). The results at $J_2=0$ 
($\rho_s = 0.165$ or $0.125$ according to whether $N=16$ is included or not) 
is 
lower than other estimates ($\rho_s \approx 0.18J_1$).\cite{chakravarty_halperin,singh_huse} The fact
that $\rho_s \rightarrow 0$ as $J_2\rightarrow J_{2c}$ is again in agreement
with expectations from the nonlinear sigma model analysis, but is of course
a trivial consequence of eq.(\ref{fsr}).

In the collinear region, there is an additional anisotropy parameter in the
effective nonlinear sigma model, and the corresponding effective parameters
therefore cannot been obtained straightforwardly from the lowest finite size
correction terms.

\subsection{Susceptibility}
An independent test of the reliability of our results
can be obtained by calculating the susceptibility $\chi$: even in an 
antiferromagnetically ordered
state, the ferromagnetic susceptibility is finite, whereas for unconventional states (e.g.
dimer or chiral), one has a spin gap and therefore a vanishing
susceptibility. The vanishing of the susceptibility can thus be associated
with the vanishing of the magnetic order parameter. Moreover, in an
antiferromagnetic state one has $\chi= \rho_s/c^2$, and we thus have a
consistency check on our calculated values for $c$ and $\rho_s$.
At fixed cluster size one has
$\chi (N) = 1/(N \Delta_T)$, where $\Delta_T$ is the excitation energy of
the lowest triplet state (which has momentum $\bbox{Q} = (\pi,\pi)$ in an
antiferromagnetic state).
An extrapolation of $\chi (N)$ to the thermodynamic limit can be performed
using the finite--size formula\cite{fisher_finite_size,runge_extrapol}
$\chi = \chi(N) - const./\sqrt{N}$, and results are shown in fig.\ref{f13}.
Again, the $N=16$ cluster behaves anomalously in that $\chi (N)$ increases
going from $N=16$ to $N=20$, whereas for bigger clusters there is the
expected decrease. In the present case, this anomaly occurs for nearly
the whole range $J_2 > 0$. Also, our result for $J_2=0$ and using
$N=20,32,36$ is $\chi=0.0671$, very close to both Monte Carlo estimates
\cite{runge_extrapol} and series expansion 
results.\cite{singh_series,singh_huse,weihong_series} We
therefore think that the $N=20,32,36$ extrapolation is the most reliable one.
It is rather pleasing to note that this independent estimate gives a
critical value for the vanishing of the susceptibility (which indicates the
disappearance of gapless magnetic excitations and therefore of long--range
antiferromagnetic order) of $J_{2c} \approx 0.42$, quite close to the
estimate we found above by considering the order parameter.

A quantitative comparison of results for the susceptibility obtained either
from the excitation gap or from the previously calculated values of $c$ and
$\rho_s$ and using $\chi = \rho_s/c^2$ reveals considerable discrepancies 
(see fig.\ref{f13}), even well away from the
``critical region'' $J_2 \approx 0.4$. The most likely explanation for
this is that our calculation of $c$ and $\rho_s$ is based on {\em
corrections} to the leading finite--size behavior, whereas $\chi$ is
obtained directly from the gap. In particular, judging from the case $J_2=0$, 
we probably underestimate the spin wave velocity by quite a bit. The direct estimate of $\chi$ is thus
expected to be more precise.

\begin{figure}
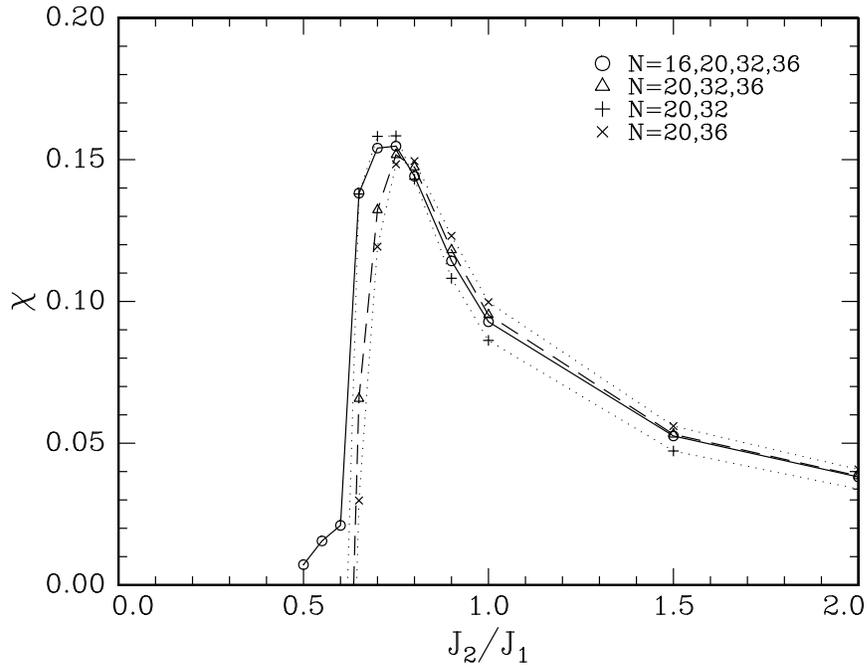

\caption{The susceptibility in the collinear region as obtained from
$\chi = 2/(N \Delta_T)$ using
different extrapolations. As discussed in the text, the $N=20,32,36$
extrapolation is expected to be the most reliable one.}
\label{f13a}
\end{figure}

An analogous calculation of the susceptibility can be performed in the
region of larger $J_2$, where the lowest excited triplet state is at
$\bbox{Q} = (\pi,0)$. In this case, because of the double degeneracy of this
state, the susceptibility is given by $\chi = 2/(N\Delta_T)$. Because of the
lower symmetry of the wavevector, the Hilbert space needed to determine the
excited state roughly double in size, and for $N=36$ has dimension 31561400.
We use the same finite--size extrapolation as before, and results obtained
for different combinations of cluster sizes are shown in fig.\ref{f13a}.
The 16 site cluster again shows rather anomalous behavior and therefore we
do not take it into account in these extrapolations. The results then
indicate a transition into a nonmagnetic ($\chi = 0$) state at $J_2/J_1
\gtrsim 0.6$, in approximate agreement with what we obtained from estimates
of the order parameter above. The decrease of $\chi$ with increasing $J_2$
is not surprising, as for large $J_2$ the model consists of two nearly
decoupled unfrustrated but interpenetrating Heisenberg models, each with
exchange constant $J_2$, and consequently one has $\chi \propto 1/J_2$.
What is a bit more surprising is the sharpness of the maximum of $\chi$
around $J_2/J_1 = 0.7$.

\subsection{Comparison with spin--wave theory}
Linear spin--wave theory (LSWT) has proven to be a surprisingly accurate
description
of the ordered state of quantum antiferromagnets even for 
spin one-half. We here compare our
numerical results with that approach. The lowest order spin--wave energies in
the antiferromagnetic and collinear state are
\begin{eqnarray}
\omega_{AF}(\bbox{k}) & = &
2 \{[1 - \alpha(1-\eta_{\bbox{k}})]^2-\gamma_{\bbox{k}}^2\}^{1/2} \;\;, \\
\omega_{coll}(\bbox{k}) & = &
\{(2 \alpha + \gamma_x)^2 - (2 \alpha \eta_{\bbox{k}} + \gamma_y
)^2 \}^{1/2} \;\;,
 \end{eqnarray}
where $\alpha = J_2/J_1$, $\gamma_{\alpha}
= \cos k_\alpha$, $\gamma_{\bbox{k}} =
(\gamma_{x}+\gamma_{y})/2$, and $\eta_{\bbox{k}} =
\gamma_{x} \gamma_{y}$.
In LSWT, the order antiferromagnetic
and collinear order parameters then are given by
\cite{chakravarty_halperin,chandra_doucot}
\begin{eqnarray}
\label{m0sw}
m_0(\bbox{Q}_0) & = &
1 - \frac{1}{4\pi^2} \int d^2k
\frac{1 - \alpha(1-\eta_{\bbox{k}})}%
{\omega_{AF}(\bbox{k})} \\
\label{m1sw}
m_0(\bbox{Q}_1) & = &
1 - \frac{1}{8\pi^2} \int d^2k
\frac{2\alpha+\gamma_x}{\omega_{coll}(\bbox{k})}
\end{eqnarray}
where the integration is over the full
first Brillouin zone. A comparison of our results with this approach is
shown in fig.\ref{f12}. For the antiferromagnetic order parameter, we
observe very satisfying agreement. What is slightly disturbing here is 
that inclusion of the next order ($1/S^2$) correction to 
eq.(\ref{m0sw})\cite{chakravarty_halperin} actually makes the agreement worse,
even for negative $J_2$ where
the next--nearest neighbor interaction stabilizes the order and spin--wave
theory therefore should be increasingly reliable. For example, for
$J_2 = -J_1$ these corrections lead to \cite{torb} $m_0(\bbox{Q}_0)= 0.775$,
whereas we find $m_0(\bbox{Q}_0)= 0.846$. To which extent higher--order spin wave theory can be used systematically even in this region thus seems unclear.
For the more interesting case of positive $J_2$, higher
corrections to spin--wave theory give more and more strongly diverging
results as $J_2 \rightarrow 0.5$, and it is not clear how any useful
information can be obtained from these higher order corrections in the
frustrated case. We therefore limit our comparison here to linear spin--wave
theory.

For the collinear state at large $J_2$, there is a similar good agreement
between spin--wave theory and our results, except for the immediate 
vicinity of the transition to the nonmagnetic state. Moreover, it appears that the prder parameter tends for large $J_2$ to a value very close or identical 
to that of the antiferromagnetic order parameter at $J_2 = 0$. This is in fact
not difficult to understand: for $J_2 \gg J_1$ our model represents two
very weakly coupled sublattices, with a strong antiferromagnetic coupling
$J_2$ within each sublattice. Consequently, the ground state wave function
is to lowest order in $J_1/J_2$ a product of the wavefunctions of
unfrustrated Heisenberg antiferromagnets on the two sublattices. We then
obtain $M_N^2(\bbox{Q}_1,J_2=\infty) = M_{N/2}^2(\bbox{Q}_0,J_2=0)/2 $, and
thus from eqs.(\ref{fsm}) and (\ref{fsm1}) we have the {\em exact} result
\begin{equation}
 \lim_{J_2/J_1 \rightarrow \infty} m_0(\bbox{Q}_1) =
  m_0(\bbox{Q}_0)|_{J_2=0} \;\;.
\end{equation}
The spin--wave results (\ref{m0sw}) and (\ref{m1sw}) as well as our numerical results do satisfy this
relation. To obtain this agreement it was however crucial to use the normalization of $M_N^2(\bbox{Q})$ shown in eq.(\ref{mq}). Using a factor $1/N^2$ instead we obtain for large $J_2$ $m_0(\bbox{Q}_1) \approx 0.4$, 
which is far too low. The reason for this is that
our extrapolation with $N=16,20,32,36$
corresponds, for large $J_2$, to a calculation on two nearly uncoupled
and unfrustrated sublattices, each with $N=8,10,16,18$. On such small
 lattices, short--range effects are obviously rather large, 
and therefore the proper normalization of $M_N^2$ is particularly important.

The ground state energy per site is given in lowest order spin--wave theory
by \cite{torb,bergomi_these}
\begin{eqnarray}
\label{es1}
e_0 & = & \frac32 (\alpha -1) + \frac{1}{8 \pi^2}
\int d^2k \; \omega_{AF}(\bbox{k})
\;\;, \\
\label{es2}
e_0 & = & - \frac32 \alpha + \frac{1}{8 \pi^2}
\int d^2k \; \omega_{coll}(\bbox{k})
\end{eqnarray}
for the antiferromagnetic and collinear state, respectively. As can be seen
in fig.\ref{f9}, these results are rather close to our finite--size
extrapolations. Nevertheless, there is a significant discrepancy: e.g. for
$J_2=0$ the spin--wave result is $e_0 = -0.6579$, compared to the presumably
best estimate from large scale Monte Carlo
calculations \cite{runge_extrapol}, $e_0 = - 0.66934$.
On the other hand, as discussed above, our finite size extrapolation gives
values very close to this. It would thus seem that, as far as the ground
state energy is concerned, finite--size extrapolation is more precise than
linear spin--wave theory.

Comparing our results for the spin--wave velocity and spin stiffness (figures
\ref{f10} and \ref{f11}) to the LSWT results $c = \sqrt{2(1-2J_2/J_1)}J_1$
and $\rho_s = (J_1-2J_2)/4$, one finds rather sizeable discrepancies, 
both for $\rho_s$ and for $c$. Nevertheless,
 the functional form for large negative $J_2$ seems to be correct.
However, here a detailed comparison seems not particularly useful as LSWT
results themselves are rather imprecise (as shown e.g. by the large
renormalization of the susceptibility at $J_2 = 0$).

\section{Summary and discussion}
In this paper we have reported detailed finite--size calculations on the
frustrated spin--1/2 antiferromagnetic Heisenberg model on the square
lattice. Using finite--size extrapolation formulae, we derived results
for a number of physical properties. The most important finding seems to
be the existence of a region of intermediate second nearest neighbor
coupling $J_2$ where no magnetic order, antiferromagnetic, collinear or
otherwise, exists. The location of the boundaries of this nonmagnetic
region depends on the cluster size involved in the estimate. For
$N=16,20,32,36$ we find the interval $0.48 \le J_2/J_1 \le 0.6$ to be
nonmagnetic, whereas with $N=20,32,36$ 
this interval is larger: $0.34 \le
J_2/J_1 \le 0.68$. Given the irregular behavior of the $N=16$ cluster we
often found above, in particular in the region of intermediate $J_2$,
the second estimate would appear to be the more reliable one.
In any case, independently of which extrapolation one prefers, there is
a nonmagnetic interval.

Beyond the existence of a nonmagnetic region, we have also obtained
quantitative estimates for a number of fundamental physical parameters
in the magnetically ordered states, antiferromagnetic for small or
negative $J_2$, collinear for large positive $J_2$. The accuracy of
these estimates can best be assessed by comparing with the unfrustrated
case $J_2 = 0$, for which case there are currently rather precise
results available, mainly from large--scale Monte Carlo calculations and
series expansions. A summary of our results, together with other recent
data, is given in table \ref{t4}. Our results for the ground state energy,
the antiferromagnetic order parameter,
and the susceptibility agree to within a percent or better,
with the best currently available numbers. 
Finally, our estimates for the
spin--wave velocity and the spin stiffness are rather imprecise. This is
certainly mainly due to the fact that these quantities are obtained from
the amplitudes of the leading correction to the asymptotic large--size
behavior of the ground state energy and the order parameter
susceptibility, and these correction are almost certainly estimated less
precisely than the leading terms.

We found it instructive to also investigate regions where magnetic order
is well--established, i.e. $J_2 \le 0$ for the antiferromagnetic case
and $J_2 \ge J_1$ for the collinear case. In these regions we find that
the finite--size formulae like (\ref{fsm}) and (\ref{fse}) provide an
excellent fit to
our numerical results. The progressive worsening of the quality of the fits
as the intermediate region is approached certainly is consistent with
the existence of a qualitatively different ground state in that region.
If on the contrary the transition between antiferromagnetic and
collinear order occurred via a strong first order transition (as
suggested by some approximate theories, see below), no such progressive
worsening is expected. We also notice in this context that the $N=16$
cluster is systematically the one exhibiting the largest deviations from
the expected behavior, probably due to its unusually high symmetry.
We thus  feel that estimates ignoring this cluster may be more
reliable.

\begin{figure}
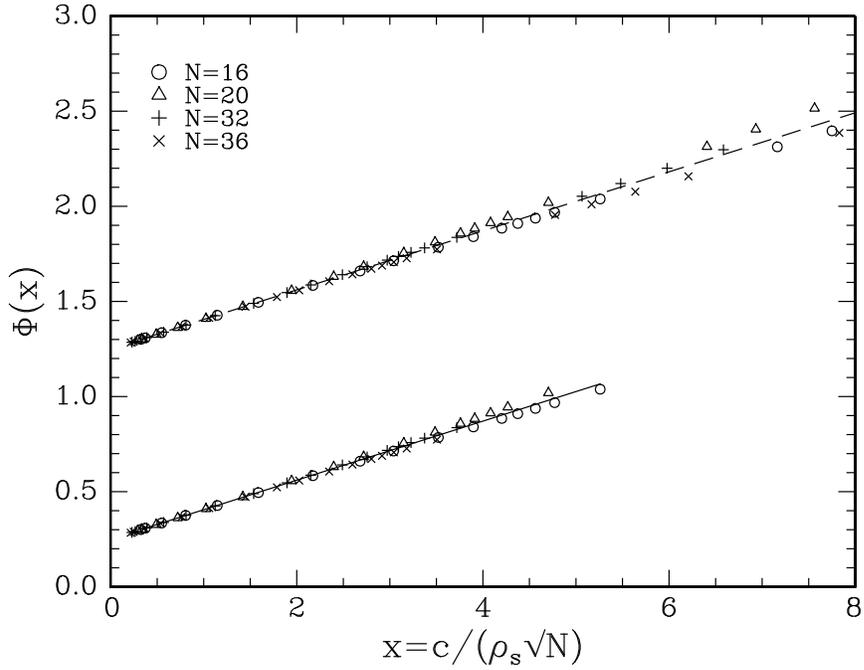

\caption[bubu]{Scaling plot of $\Phi(x)
=M_N^2(\protect{\bbox{Q}}_0)/m_0(\protect{\bbox{Q}}_0)^2$ as a function of
the
variable $x = c/(\rho_s \sqrt{N})$, using the $N=20,32,36$ (lower curve) and
$N=16,20,32,36$ (upper curve) extrapolations for $c$ and $\rho_s$.
For clarity, data for the $N=16,20,32,36$ extrapolation are shifted upward
by 3 units.
The straight lines represent the spin wave result $\Phi (x)=
(1  + 0.6208 x)/4$ }
\label{f14}
\end{figure}

Another way to assess the consistency of the finite--size extrapolations we
are
using is to verify the underlying scaling hypothesis via a ``scaling
plot''. The fundamental constants $c$ and $\rho_s$ of the nonlinear sigma
model define a length scale $c/\rho_s$, and if finite size scaling is
verified one therefore expects all finite size corrections to be universal
functions of the variable $x = c/(\rho_s \sqrt{N})$. In particular, for the
order parameter susceptibility we expect
\begin{equation}
M_N^2(\bbox{Q}_0) = m_0(\bbox{Q}_0)^2 \Phi(x) \;\;.
\end{equation}
Combining the second form of eq.(\ref{fsm}), eq.(\ref{fsr}),
and this definition,
the small--$x$ expansion of the
scaling function is $\Phi(x) =  (1  + 0.6208 x)/4$. Plots of our results
for $M_N^2(\bbox{Q}_0)$ as a function of the scaling vaiable $x$ are shown
in fig.\ref{f14}. One sees that for the $N=20,32,36$ extrapolation the plot
is nearly perfect in that nearly all data points are collapsed onto a single
curve. The only points that show a significant deviation are those obtained
for $N=16$ close to the phase transition to the nonmagnetic state. This of
course is nothing but a manifestation of the anomalous behavior of this
cluster already found previously. The behavior for the $N=16,20,32,36$
extrapolation is clearly less satisfying. A similar scaling plot for the
ground state energy produces even better results, due to the better
convergence of the corresponding finite--size formula (\ref{fse}).

\begin{figure}
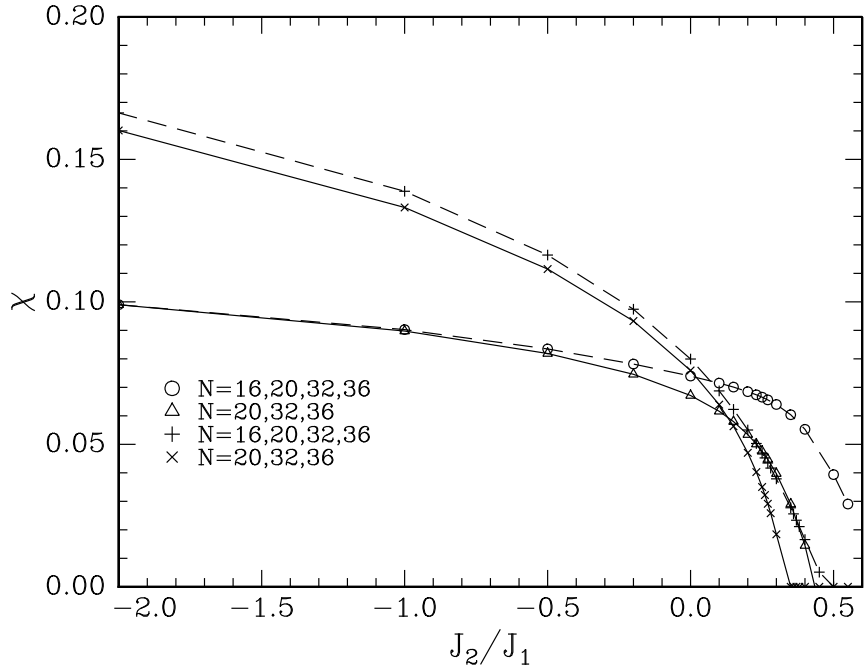

\caption{The susceptibility in the N\'eel region as obtained from $\chi =
1/(N \Delta_T)$
(circles and triangles) and from $\chi =\rho_s/c^2$ (crosses) using
different extrapolations. As discussed in the text, the $N=20,32,36$
extrapolation is expected to be the most reliable one.}
\label{f13}
\end{figure}

\begin{figure}
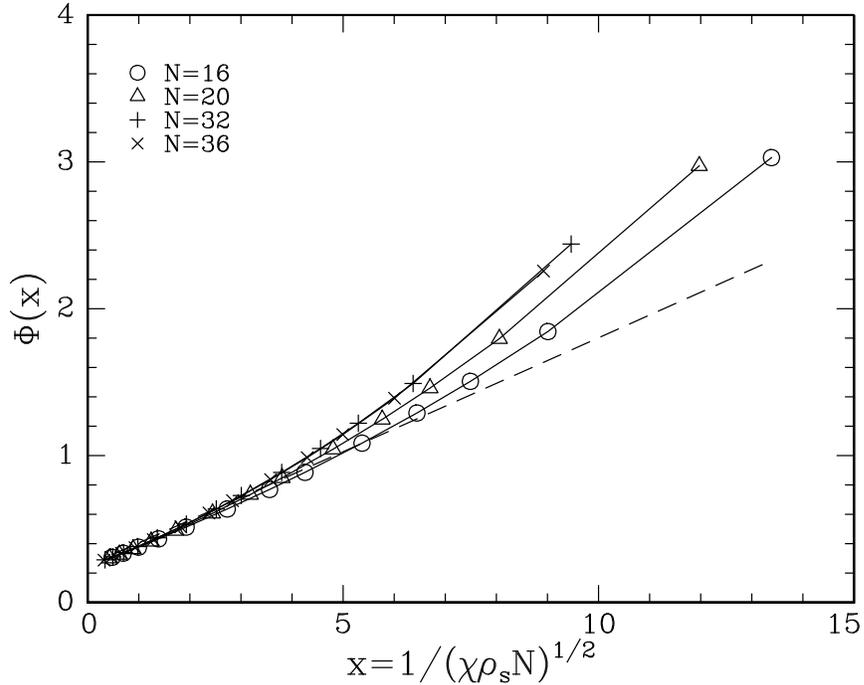

\caption[bubu]{Scaling plot of $\Phi(x)
=M_N^2(\protect{\bbox{Q}}_0)/m_0(\protect{\bbox{Q}}_0)^2$ as a function of
the
variable $x = 1/(\chi \rho_s N)^{1/2}$, using the $N=20,32,36$ results for
$\chi$ (see fig.\ref{f13}) and the $N=20,32,36$  extrapolation for
$\rho_s$. The dashed line represents the spin wave result $\Phi (x)=
(1  + 0.6208 x)/4$ }
\label{f16}
\end{figure}

A scaling plot like fig.\ref{f14} permits to assess the consistency of data
obtained for clusters of different sizes, however, the form of the scaling
function itself is obviously less significant as the coefficients $c$ and
$\rho_s$ entering the definition of the scaling variable $x$ are calculated
assuming finite--size formulae like (\ref{fsm}) and (\ref{fse}) to be valid,
i.e. implicitly {\em assuming} the form $\Phi(x) =  (1  + 0.6208 x)/4$. An
{\em independent} estimate of $\Phi$ can in principle be obtained using
independent estimates for $c$ and $\rho_s$. We do not have currently such an
estimate for $\rho_s$, however we can use our independent results for the
susceptibility (fig.\ref{f13}) to rewrite the scaling variable as
$x=1/\sqrt{\chi \rho_s N}$. The plot obtained using estimates for
$\rho_s$ and $\chi$ from $N=20,32,36$ is shown in fig.\ref{f16}.
The collapse of data obtained for different sizes and values of $J_2$  is not as satisfactory as
in the previous case, however, this is certainly related to the fact that
here we use a second independently estimated quantity, namely $\chi$. Still,
for $x \lesssim 5$, the collapse is rather good, showing the consistency of
our analysis in this region. For the larger clusters, this region
corresponds to $J_2/J_1 \le 0.25$, i.e. it extends rather close to the
transition which occurs at $J_2/J_1 \approx 0.34$. For small $x$ the
calculated scaling function essentially agrees with the spin--wave results
shown by the dashed line in fig.\ref{f16}. For $x \gtrsim 5$, there are
discrepancies between results obtained from $M_N^2(\bbox{Q}_0)$ for
different $N$. This probably indicates that at least for the smaller
clusters, finite size effects become so important that it is no more
sufficient to include the lowest order finite size corrections only.
The fact that the numerically found scaling function is larger than the
spin--wave approximation is not entirely unexpected: in fact, for large $x$,
i.e. in the critical region, one would expect $\Phi(x) \propto x^{1+\eta}$,
where $\eta$ is the correlation exponent of the three--dimensional
Heisenberg model. However, we doubt that what we observe in fig.\ref{f16}
is actually a critical effect. First, the numerical value \cite{peczak} 
of $\eta=2-(\gamma/\nu)$ is very
small: $\eta \approx 0.03$, and one thus expects an extremely smooth
crossover. Moreover, in fig.\ref{f16} we have used the independently
calculated susceptibility (see fig(\ref{f13}) which goes to zero only at
$J_2/J_1 \approx 0.42$, rather than at $J_2/J_1 \approx 0.34$ where our
estimated staggered magnetization vanishes. Consequently, the abscissae of
the data points in fig.\ref{f16} are underestimated, i.e. the data in
fig.\ref{f16} overestimate the true $\Phi(x)$.

The $J_1-J_2$ model we have studied here has been investigated previously by
number of techniques. Previous finite--size studies
\cite{dagotto_moreo,figueirido_16site}
found some indication of an intermediate phase without magnetic order,
however due to the limitation to $N=16$ and $20$ only, it was impossible to
make extrapolations to the thermodynamic limit and to arrive at quantitative
statements. Our own previous study \cite{schulz_ziman}, using $N=16$ and
$36$, produced results very similar to our current best estimates. However,
due to the larger number of clusters we now use (and due to the
possibility to ignore the anomalous $N=16$ cluster), we feel that our
conclusions are considerably more reliable.

\begin{figure}
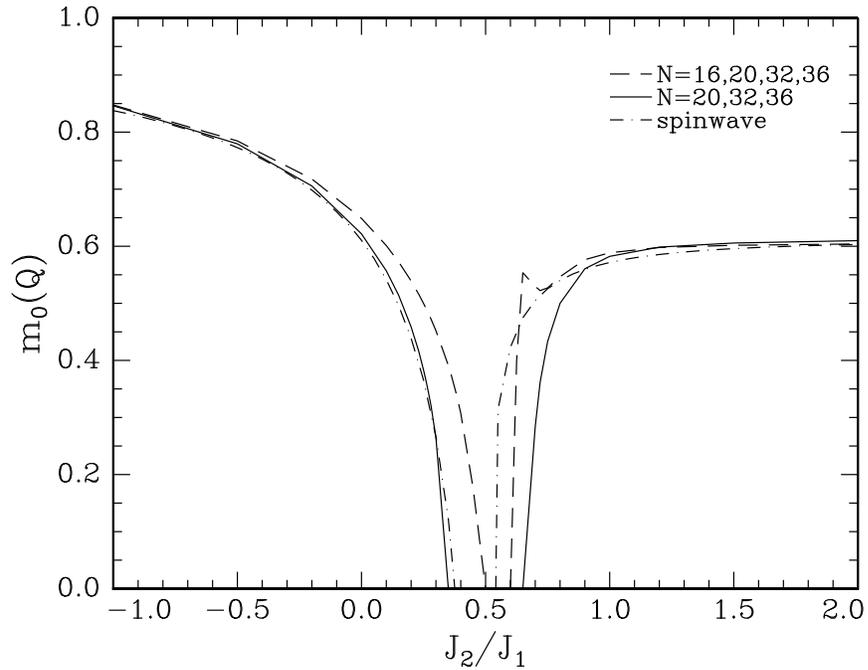

\caption{Comparison of our finite size fits for the antiferromagnetic and
collinear order parameters (left and right curves, respectively) with
linear spin wave theory.}
\label{f12}
\end{figure}

Lowest order spin--wave theory \cite{chandra_doucot}
produces a phase diagram very similar to ours (see fig.\ref{f12}). On the
other hand, higher order (in $1/S$) calculations do not seem to be very
useful, due to increasingly strong singularities at $J_2= J_1/2$. It has been
attempted to include higher order corrections using a selfconsistently
modified spin--wave theory.\cite{takahashi_spinwave,nishimori_spinwave}
These calculations as well as the closely related Schwinger boson
approach \cite{mila_bruder} produce a first order transition between N\'eel
and collinear state. 
A combination of Schwinger boson results and a 
renormalization group calculation\cite{einarsson_seff} 
gives on the other hand a 
second order transition from the N\'eel state
to a magnetically disordered state, at $J_{2c}/J_1 = 0.15$.\cite{einarsson}
However, the applicability of these approaches to an $S=1/2$ system is
hard to judge, mainly due to the absence of a small parameter that would make 
a systematic expansion possible. 

Quantum Monte Carlo methods are plagued with the sign problem for
frustrated
spin systems. Nevertheless, conclusions very similar to the modified spin
wave calculations have
been reached recently using a quantum Monte Carlo method.\cite{nakamura_mc}
However, these results have rather large error bars and in some cases, in
particular in the region of intermediate $J_2$, are in disagreement with our
present exact results. The validity of these results thus appears doubtful
to us.

Another approach has been via series expansion methods around a lattice
covered by isolated dimers.\cite{gelfand_series} Expanding around a
columnar arrangement of dimers, these authors find a phase diagram very
similar to ours, at least as far as magnetic order is concerned. However,
these results are not without ambiguity: expanding around a staggered dimer
arrangement, there appears to be a first order transition between N\'eel and
collinear states. The results of this method thus appear to be biased by
the starting point of the expansion.

The most obvious question left open by the present study is the nature
of the ground state in the intermediate nonmagnetic region. Work
extending our previous analysis \cite{schulz_ziman} is in progress and
will be reported in a subsequent publication. It would also be interesting
to investigate dynamical correlations functions, in particular in the
vicinity of the critical point of the N\'eel state, $J_{2c} = 0.34$. One thus
might gain additional insight into dynamical properties at a quantum
critical point.\cite{sachdev_ye,chubukov_critical} Finally, one might also
try to extend
the size of the available clusters, in order to achieve better accuracy and
reliability. The next useful cluster has 40 sites, and should be tractable
in the near future. However, the next step then would be a cluster of 52
sites which would require computational means both in memory size and speed
three or four orders of magnitude more powerful than what is currently
available. A viable alternative to increase the size of the tractable
clusters might be to combine the exact solution of moderately big clusters
with Monte Carlo type approaches.

\acknowledgements We thank T. Einarsson for a number of helpful comments.
The numerical calculations reported here were possible thanks to computing
time made available by CCVR, Palaiseau (France) and IDRIS, Orsay (France).
The staff at these computing centers, in particular M.A. Foujols, provided
invaluable help with various computing problems. D. P.
acknowledges support from the CEE Human Capital and Mobility
Program under grant no. CHRX-CT93-0332.
H.J.S. was in part supported by CEE research contract no. CII/0568.


\begin{thebibliography}{10}
\bibitem{rem2}
On a recently available Cray C98, CPU times cited here typically are decreased
  by a factor six.
\bibitem{com_m0} Using a prefactor $1/N^2$ instead of $1/(N(N+2))$ in eq.(\ref{mq}), we
find $m_0(\bbox{Q}_0)=0.554$ and $m_0(\bbox{Q}_0)=0.532$ using $N=16,20,32,36$ and
$N=20,32,36$, respectively (see also ref.\onlinecite{schulz_ziman}).
The correct prefactor thus considerably improves the precision of the 
estimates.
\bibitem{erratum}
Note that this is the correct form, as was used in 
ref.\onlinecite{schulz_ziman}. In
   ref.\onlinecite{neuberger_ziman} the relation was misquoted.
\end{thebibliography}

\onecolumn
\narrowtext
\begin{table}
\caption{The number of states in the Hilbert space ($n_h$) and
the number of nonzero off--diagonal matrix elements ($n_e$)
for the clusters used in this paper. The numbers are for states
in the $A_1$ representation ($A$ representation for $N=20$)
at momentum $\protect \bbox{Q} =0$.
\label{t1}}
\begin{tabular}{lrr}
N       &   $n_h$              & $n_e$ \\
\tableline
16 & 107 & 3664 \\
20 & 1,321 & 55,660 \\
32 & 1,184,480 & 78,251,988 \\
36 & 15,804,956 & 1,170,496,152 \\
\end{tabular}
\end{table}

\mediumtext
\begin{table}
\caption{The ground state energy per site for different clusters and
different values of $J_2$ ($J_1$ is normalized to unity). Where the ground
state representation changes with increasing $J_2$ ($N=20,36$), the energies
of both relevant representations are given.
Boldface indicates the approximate location of changes in the ground state
symmetry.
\label{t2}}
\begin{tabular}{dcccccc}
$J_2$ & 16 & 20($A$) & 20($B$) & 32 & 36($A_1$) & 36($B_1$)\\
\tableline
-1.00 &  -1.16457   &  -1.15103 & -0.801770   &  -1.13251  & -1.12922   &       \\
-0.50 &  -0.927249  & -0.915408 & -0.648275   & -0.900134 & -0.897626   &      \\
 0.00 &  -0.701780  & -0.690808 & -0.519508   & -0.680179 & -0.678872   & -0.603912 \\
 0.10 &  -0.659817  & -0.648444 & -0.501316   & -0.639048 & -0.638096   &      \\
 0.20 &  -0.619874  & -0.607519 & -0.487925   & -0.599542 & -0.599046   &      \\
 0.30 &  -0.582984  & -0.568545 & -0.479923   & -0.562283 & -0.562459   &      \\
 0.40 &  -0.551147  & -0.532381 & -0.476480   & -0.528379 & -0.529745   &      \\
 0.50 &  -0.528620  & -0.500615 & -0.476624   & -0.500096 & -0.503810   &  -0.493941   \\
 0.55 &  -0.523594  & {\bf -0.487338} & -0.478122   & -0.489517 & -0.495178   & -0.490396   \\
 0.60 &  -0.525896  & -0.491633 & {\bf -0.491816}   & -0.484599 & -0.493239   &  -0.492267   \\
 0.65 &  -0.539382  & -0.516444 & -0.517029   & -0.502147 & {\bf -0.506588}   &  -0.506582   \\
 0.70 &  -0.563858  & -0.543309 & -0.545677   & -0.527741 & -0.529951   & {\bf -0.530001}   \\
 0.80 &  -0.627335  & -0.600092 & -0.609595   & -0.586871 & -0.585428   &  -0.586487   \\
 0.90 &  -0.696866  & -0.659162 & -0.677703   & -0.651509 & -0.645445   & -0.649052   \\
 1.00 &  -0.768468  & -0.719583 & -0.747576   & -0.718414 & -0.707495   & -0.714360    \\
 1.20 &  -0.914286  & -0.842827 & -0.88967    & -0.854910 &             & -0.848364  \\
 1.50 &  -1.13578   &  -1.03098 & -1.10536    &  -1.06229 &             & -1.05268      \\
 2.00 &  -1.50771   &  -1.34863 & -1.46744    &  -1.41044 &             & -1.39633       \\
\end{tabular}
\end{table}
\mediumtext
\begin{table}
\caption{The normalized susceptibility (eq.(\protect\ref{mq})) at
$\protect \bbox{Q} = (\pi,\pi)$ and
$\protect \bbox{Q} = (\pi,0)$ for different clusters and different values of
$J_2/J_1$. \label{t3}}
\begin{tabular}{dcccccccc}
& \multicolumn{4}{c}{$M(\pi,\pi)$} & \multicolumn{4}{c}{$M(\pi,0)$} \\
 $J_2/J_1$ & 16 & 20 & 32 & 36 & 16 & 20 & 32 & 36 \\
\tableline
 -1.00  & 0.26924  & 0.26002  & 0.24296  & 0.23943  & 0.02778 & 0.02273  & 0.01470  & 0.01316 \\
 -0.50  & 0.26297  & 0.25221  & 0.23131  & 0.22660  & 0.02780 & 0.02273  & 0.01471  & 0.01316 \\
  0.00  & 0.24580  & 0.23430  & 0.20621  & 0.19879  & 0.02789 & 0.02278  & 0.01476  & 0.01322 \\
  0.10  & 0.23853  & 0.22785  & 0.19745  & 0.18893  & 0.02798 & 0.02284  & 0.01480  & 0.01326 \\
  0.20  & 0.22811  & 0.21949  & 0.18616  & 0.17601  & 0.02818 & 0.02292  & 0.01489  & 0.01335 \\
  0.30  & 0.21212  & 0.20816  & 0.17090  & 0.15800  & 0.02868 & 0.02309  & 0.01506  & 0.01354 \\
  0.40  & 0.18589  & 0.19193  & 0.14887  & 0.13109  & 0.03031 & 0.02348  & 0.01545  & 0.01404 \\
  0.50  & 0.14236  & 0.16693  & 0.11487  & 0.09236  & 0.03709 & 0.02452  & 0.01669  & 0.01594 \\
  0.55  & 0.11276  & 0.14834  & 0.09165  & 0.07062  & 0.04771 & 0.02621  & 0.01880  & 0.01965 \\
  0.60  & 0.07819  & 0.02915  & 0.05113  & 0.04378  & 0.07154 & 0.11508  & 0.04627  & 0.03822 \\
  0.65  & 0.04290  & 0.02015  & 0.01692  & 0.01954  & 0.10897 & 0.12615  & 0.10333  & 0.08167 \\
  0.70  & 0.02092  & 0.01303  & 0.01161  & 0.01232  & 0.13598 & 0.13461  & 0.11321  & 0.10006 \\
  0.80  & 0.00721  & 0.00561  & 0.00520  & 0.00611  & 0.15407 & 0.14383  & 0.12265  & 0.11370 \\
  0.90  & 0.00374  & 0.00302  & 0.00251  & 0.00314  & 0.15930 & 0.14759  & 0.12616  & 0.11925 \\
  1.00  & 0.00236  & 0.00193  & 0.00147  & 0.00183  & 0.16164 & 0.14944  & 0.12759  & 0.12183 \\
  1.20  & 0.00126  & 0.00103  & 0.00072  & 0.00088  & 0.16379 & 0.15118  & 0.12876  & 0.12418 \\
  1.50  & 0.00067  & 0.00055  & 0.00037  & 0.00044  & 0.16507 & 0.15227  & 0.12939  & 0.12553 \\
  2.00  & 0.00033  & 0.00027  & 0.00018  & 0.00021  & 0.16586 & 0.15295  & 0.12977  & 0.12637 \\
\end{tabular}
\end{table}

\narrowtext
\begin{table}
\caption[bb]{Comparison of our results at $J_2=0$ obtained from the
$N=16,20,32,36$ and $N= 20,32,36$ extrapolations with previous estimates
from series expansions and quantum Monte Carlo calculations. A more complete
compilation of previous results can be found in review 
articles.\cite{manousakis_review,barnes_review}
\label{t4}}
\begin{tabular}{lccc}
       & $e_0$ & $m_0(\bbox{Q}_0)$  & $\chi$ \\
\tableline
$N=16,20,32,36$  & -0.6688 & 0.649 & 0.0740 \\
$N= 20,32,36$    & -0.6702 & 0.622 & 0.0671 \\
series expansions \tablenote{See refs. \onlinecite{singh_series} and
\onlinecite{weihong_series}.}& -0.6696 & 0.614 & 0.0659 \\
quantum Monte Carlo \tablenote{See refs. \onlinecite{runge_mag} and
\onlinecite{runge_extrapol}.}&  -0.6693 & 0.615 & 0.0669  \\
\end{tabular} \end{table}

\renewcommand{\baselinestretch}{1.0}

\begin{figure}
\caption{The clusters used in this paper.}
\label{f1}
\end{figure}

\newpage

\newpage

\newpage

\newpage
\begin{figure}
\caption{The collinear magnetization $m_0(\protect{\bbox{Q}}_1)$ as a
function of $J_2/J_1$ using different combinations of clusters (a). In (b)
the ``critical'' region $0.5\le J_2 \le 1.0$ is shown enlarged.}
\label{f7}
\end{figure}

\newpage

\newpage

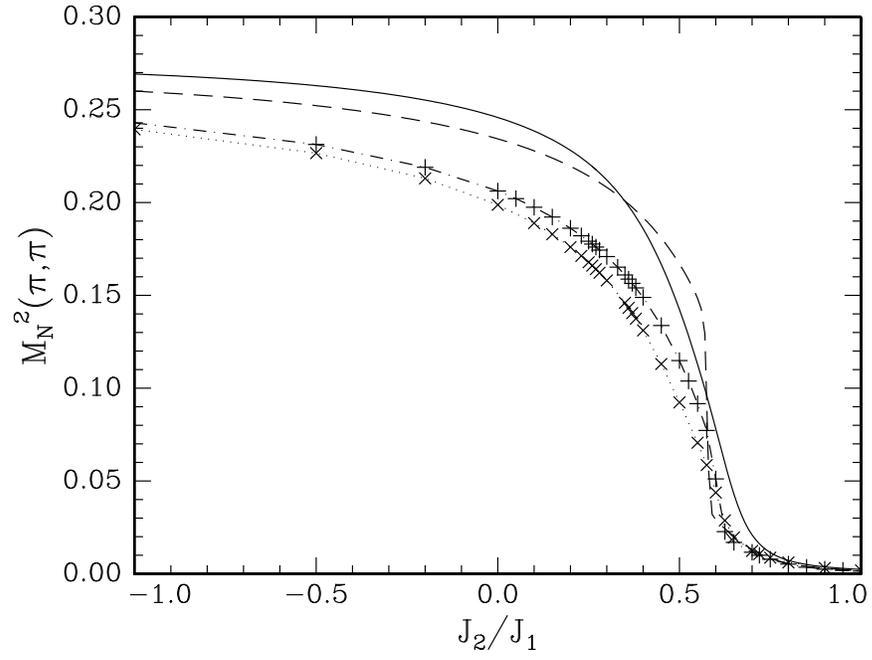
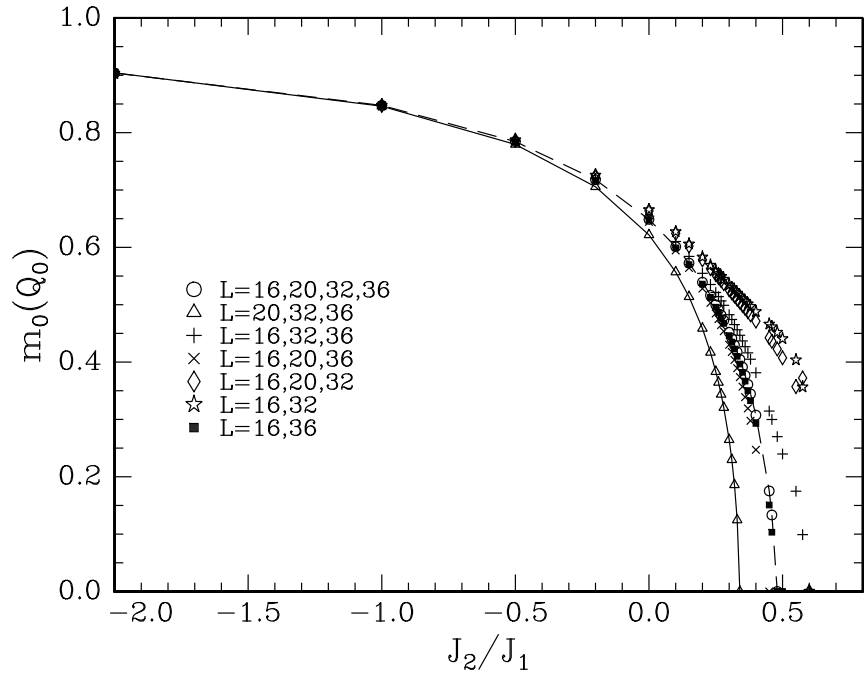
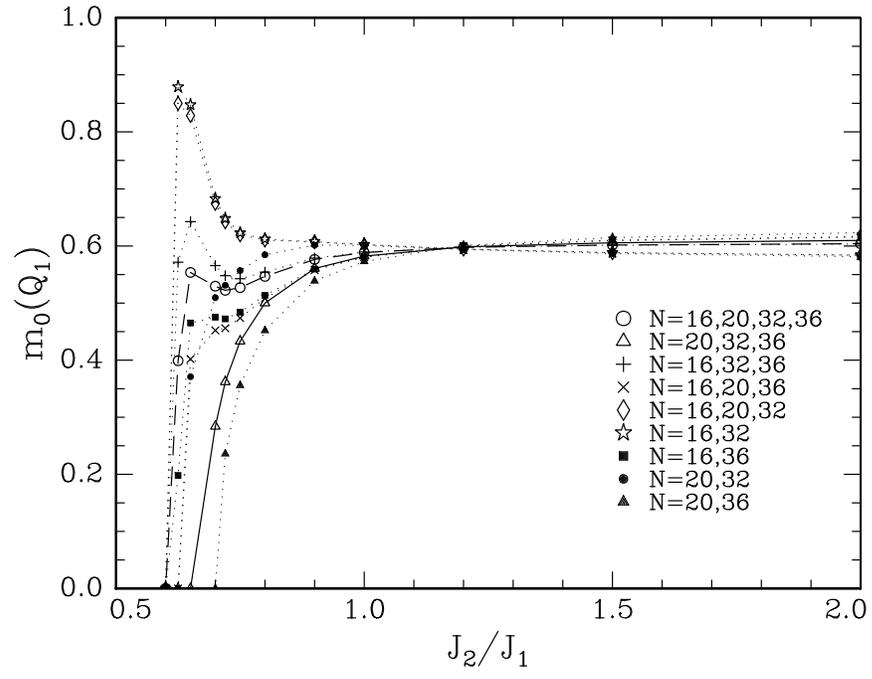
\end{document}